\documentclass[12pt]{article}

\usepackage[utf8]{inputenc}
\usepackage{alphabeta}
\usepackage[a4paper,margin=1in,footskip=0.2in]{geometry}
\usepackage[usenames,dvipsnames]{color}
\usepackage{authblk}
\usepackage{ragged2e}
\usepackage{amsmath}
\usepackage{longtable}
\usepackage{mathtools}
\usepackage{amsfonts}
\usepackage{booktabs}
\usepackage{siunitx}
\usepackage{url}
\usepackage{array}
\usepackage{bbding}
\usepackage{amssymb}
\usepackage{eucal}
\usepackage{graphicx}
\usepackage{multicol}
\usepackage{adjustbox}
\usepackage{subcaption}
\usepackage{booktabs}
\usepackage{float}
\usepackage{appendix}
\numberwithin{equation}{section}
\usepackage{wasysym}
\usepackage{rotating}
\usepackage{listings}
\usepackage{tikz}
\usepackage{graphicx} 
\usetikzlibrary{calc}
\usetikzlibrary{matrix}
\usetikzlibrary{positioning}
\usepackage{color}
\usepackage{setspace}
\usepackage{hyperref} 
\usepackage{natbib}
\usepackage{appendix}

\begin{document}
\begin{titlepage}
\title{
 The echo chamber effect resounds on financial markets: a social media alert system for meme stocks
\\
 \vspace{0.5cm}
}

\author[a,\footnote{
E-mail addresses: ilaria.gianstefani@imtlucca.it (I. Gianstefani),
luigi.longo@imtlucca.it (L. Longo)
  massimo.riccaboni@imtlucca.it (M. Riccaboni). 
}]{Ilaria Gianstefani}
\author[a]{Luigi Longo}

\author[a]{Massimo Riccaboni}
\affil[a]{
\begin{footnotesize}Institute for Advanced Studies Lucca, Piazza S.Francesco, 19, 55100 Lucca (LU) 
\end{footnotesize}
}

\vspace{-1cm}
\begin{normalsize}
\date{\today}
\end{normalsize}

\vspace{-1cm}
\maketitle
\begin{abstract}
The short squeeze of Gamestop (GME) has revealed to the world how retail investors pooling through social media can severely impact financial markets. In this paper, we devise an early warning signal to detect suspicious users' social network activity, which might affect the financial market stability. We apply our approach to the subreddit \textit{r/WallStreetBets}, selecting two meme stocks (GME and AMC) and two non-meme stocks (AAPL and MSFT) as case studies. The alert system is structured in two stpng; the first one is based on extraordinary activity on the social network, while the second aims at identifying whether the movement seeks to coordinate the users to a bulk action. We run an event study analysis to see the reaction of the financial markets when the alert system catches social network turmoil. A regression analysis witnesses the discrepancy between the meme and non-meme stocks in how the social networks might affect the trend on the financial market. 

\bigskip
\end{abstract}
\setcounter{page}{0}
\thispagestyle{empty}
\end{titlepage}
\pagebreak \newpage


\newpage

\section{Introduction}
Retail investors have always been considered as noise traders in the financial and market microstructure literature: their choices are not driven by the knowledge of the fundamentals of a stock or any sophisticated analysis of the market, but they are guided mainly by their emotional and irrational beliefs. Noise traders market impact has always been considered negligible compared to the influence of large players (such as investment banks and hedge funds). However, this picture of nonthreatening amateur investors seems outdated: the progressive diffusion of social media combined with low cost trading platforms are making investment strategies more and more widespread. The impact they have on the financial market is anything but negligible. The most striking episode is the short-squeeze that some retail investors triggered on GameStop stock (ticker symbol on NYSE:  GME) by coordinating themselves on Reddit, a micro-blogging social network \citep{anand2021}.\\ 
Reddit is a website\footnote{In the top-10 of the most visited websites in the US according to various websites such as \url{https://www.alexa.com/topsites/countries/US} and \url{https://www.semrush.com/blog/most-visited-websites/} } composed of user-generated content and related discussions. The site's content is divided into forums, communities known as "subreddits", which deal with a specific topic. As a network of communities, Reddit's core content consists of posts (submissions) from its users. Users can comment on others' posts to continue the conversation, and they can collect positive or negative votes (score). The number of upvotes or downvotes determines the posts' visibility on the site; the more popular the content is, the higher the number of people it is displayed.\\
One of the most popular and active subreddit is \textit{r/WallStreetBets}, a community focused on financial markets and stock trading. In this community, users boast very aggressive trading strategies and what they did at the end of January 2021 with GameStop is no exception. GameStop (GME) is the world's largest American retailer of video games and accessories. The market for physical video games started its decay in 2017 due to the downloadable version of many games and services offered by the main consoles. GameStop started facing a sharp decline in sales that determined a share-price dropping in the financial market. The COVID-19 pandemic was a severe beating for the company, and its fundamentals faced another shock. The stock price decline led many institutional investors to sell the stock short. Conversely, some retail investors, considering the stock undervalued, went against the trend of the big players. In January 2021, a coordinated effort orchestrated by the community of the subreddit \\\textit{r/WallStreetBets} surged the price of GME \citep{StaffReportUS21}.\\
Apart from what happened on the financial market, the squeeze of the price and the consequent losses faced by the short-sellers, the high volatility of traded volumes and the liquidity issues, the most striking part of this episode is to comprehend how an apparently harmless group of noise traders were able to provoke such a substantial effect on a market usually dominated by the big players.\\
Stylizing the phenomenon can be helpful to detect eventual anomalies based on indicators of coordination. This is relevant in a perspective of policy-making to prevent this kind of shock from happening or at least to tackle the harmful effects.\\ 
Besides the specific case of GameStop, which is dramatic in terms of magnitude and subsequent effects, the interest in how social media networks impact the financial price formation of meme stocks is gaining growing attention \citep{Pedersen2021,costola2021}. Furthermore, an ever-increasing decentralization of the financial system and the ease to access it via user-friendly online trading platforms are potential destabilizers for the financial ecosystem\footnote{See for example: https://www.risk.net/investing/7810026/aqr-quant-on-the-network-effects-behind-gamestop-frenzy}. The fintech (r)evolution and the capillary diffusion of the online social network, which allow us to receive the up-to-date information as they happen, lay the foundations for a reinterpretation of market manipulation steering through social networks.
Hence, here is the urgent need to create an alert system to pinpoint episodes of misconduct behaviors in online social network that can result in potential market manipulation. 
With misconduct behaviors, in this paper we refer to an inappropriate activity at the social network level that might potentially translate into market manipulation, a deliberate attempt to interfere with the free and fair operation of the market.\\
Although there is no regulation legislating on the relationship between social media coordination and the financial market, the U.S. securities and exchange commission (SEC) defines as market manipulation\footnote{https://www.investor.gov/introduction-investing/investing-basics/glossary/market-manipulation} all the actions where someone artificially affects the supply or demand for a security.\\ Social network variables on mass coordination are valuable tools for building nowcasting systems and scheduling real-time interventions to ensure stability.
Our paper contributes to the extant literature by designing an alert system to detect potential misconducting behaviors or suspicious activity on the social network to eventually prevent the harmful coordination from creating instability in financial markets. The methodology relies on social listening and social network analysis to identify the red light days to monitor.
The rest of the paper proceeds as follows: Section 2 reviews the literature on the topic, Section 3 presents the data and their features, Section 4 describes the alert system we develop, and Section 5 introduces the empirical results. Finally, Section 6 concludes.\\

\section{Literature Review}
Digital and online social network revolutions are deeply affecting the functioning of financial markets. One manifestation of this revolution is the rising importance of retail traders. In the classical market microstructure models \citep{GlostenMilgrom1985,Kyle85}, noise traders are considered as a residual category because of their randomness in the trades and are usually ignored in the price formation process because of their irrational impact on the market (which temporary makes the price to diverge from the fundamental value) is predominated and counterbalanced by rational agents on the market.\\
One of the first models to recognize the relevance of noise traders in the financial ecosystem is the one described by \citet{DeLongEtAl1990} where the authors model them as irrational traders with erroneous stochastic beliefs whose \textit{unpredictability creates a risk in the price of an asset. As a result, prices can diverge significantly from fundamental values even without fundamental risk}. This paper was highly enlightened and forward-looking as it perceived how the agents, so far targeted as irrelevant, are effectively impacting the asset price formation in certain circumstances.\\
Thanks to user-friendly, low-cost online trading platforms, the widespread diffusion of social media and the easiness of accessing financial markets have significantly transformed the market dynamics. 
As pointed out by \citet{ZheludevSmithAste2014}, the proliferation of the internet has improved our ability to access information in real-time, and in particular, the diffusion of social media allows us to get in contact with the moods, thoughts, and opinions of a large part of the world's traders in an aggregated and real-time manner. Many are the researches aim to analyze the impact of social media on the financial markets both under a perspective of the volume of the social activity, the search engine traffic, and the prevailing sentiment of the agents.
A consistent branch of the literature focuses, using various techniques, on the impact of social media activity on the financial markets \citep{MaoEtAL2012,BordinoEtAl2012,RuizEtAl2012,PreisEtAl2013} . Most of the results presented so far are based on empirical extrapolation. \\
In \citet{Renault2018}, the author focuses on a specific type of market manipulation: the pump-and-dump scheme. He finds that an abnormal activity on social media about a specific stock is associated with a sharp increase in volume and price on the day of the event, while the price presents a reversal over the following trading week.
\citet{Pedersen2021} proposes a new model that revolutionizes the vision of the so-called noise traders. He describes how investment strategies propagate in a social network and how they affect the market. Four typologies of investors are considered: besides the classic prototypes of rational short- (who tries to predict the sentiment changes among naive investors based on social network dynamics) and long-term investor (focused on the fundamental value of assets), it portrays two new types of agents: the 'fanatics' (investors with a stubborn view that can influence many people thanks to their popularity on social media) and naive investors (agents learning and relying on investment strategies proposed on social networks). The model explains the belief formation process on the social network and how it affects the fluctuations of prices and trading volumes on the financial market. 
Modern social media, such as Reddit, allows envisioning (and downloading) the data generating process that leads to the coordination of the agents on the network and analyzes the underlying forces behind the event. To the extent of our knowledge, there are no works devoted to studying the network evolution and consequent market impact. Suppose social network activity does generate a force that drives the financial market dynamics. In such a case, the GameStop saga can be an exciting case study to understand the network indicators to monitor to prevent extreme phenomena like the one that happened at the end of January 2021. Hence, as recommended by \cite{Pedersen2021} in the conclusion of his paper, the availability of data from social media might open \textit{'new research possibilities to test model's prediction using data on networks and market behaviors'}.\\
\citet{Dim2021} analyzes the implications on the financial markets of \textit{non-preofessional social media investment analysts} that publish investment strategies shaping the views and actions of many retail investors. This study highlights how the interplay of social media and retail trading poses new challenges for financial markets and regulators, which requires a deeper understanding of belief formation on social media.\\
Our paper works towards defining some indicators and parameters to monitor on the social network to detect extreme situations that might affects the financial market stability. Inspired by the setting proposed in \citet{costola2021}, our alert system has two consecutive red flags: if the first one, based on extraordinary activity on the social network, activates, we start monitoring the structure of the user social network.\\
In the network analysis, distinguishing the various roles the users can play is crucial. Many works are devoted to this categorization \citep{RiosEtAl2019,ChoiEtAl2015,ThukralEtAl2018}. 
A special role is played by the influencers, aka the leaders, or to use a term proposed \cite{Pedersen2021}, the "fanatics". We track the users' activity within the network, catch the agents distinguishing for their ability to be central and vocal in the network by proposing relevant contributions.\\
Before delving into the core of our paper, we cross-reference all the works dealing with the GameStop case, the triggering factor of this piece of literature devoted to creating an alert system to prevent the dysfunctional social network activity destabilizing the financial market.
Investment recommendation \citep{BradleyEtAl2021}, social network activity volume and tone (obtained with sentiment analysis) \citep{LongEtAl2021,UmarEtAl2021} influence the GME returns and traded volume on the financial market. Also, the Google trend researches with keywords related to the event are positively correlated with the financial GameStop performance \citep{Klein2021,VasileiouEtAl2021}. \citet{Hasso2021} profile the agents participating in the frenzy and describe their average performance; they have proven to be relatively inexperienced and unsophisticated \citep{EatonEtAl2021}.  \citet{EatonEtAl2021} also infers that a large portion of agents acting on the subreddit \textit{r/WallStreetBets} uses Robinhood as a trading app. Robinhood is a zero commission, no account minimum, and an easy-to-use interface trading app widely used among young investors. During the most turbulent period of GME frenzy, the trading app went down or malfunctioned several times, avoiding investors from acting on the financial market and loosing the best moments to trade. Many were the complains about this malfunction reported on a website \textit{DownDetector.com}\footnote{https://downdetector.com/}, an online platform that provides users with information about the status of various websites and services based upon user outage reports.\\
The disruptive effect would have been milder if only the financial market presented more substantial barriers to entry.
Finally, in \citet{FusariEtAl2021}, they report as a case study the extreme event of GME, demonstrating that they can predict the bubble using a model based on options.\\

\section{Data}
Our analysis is run on four NYSE-listed stocks in the period October 2020 - June 2021 for the following stocks (NYSE ticker is reported in parenthesis): GameStop (GME), American Multi-Cinema Entertainment (AMC), Apple Inc. (AAPL) and Microsoft Corporation (MSFT). For each stock, we download the financial daily data on price and trading volume and all the post on the social network Reddit containing the ticker. GME and AMC are two examples of meme stocks, meaning stock that gains popularity among retail investors through social media. As reported in \citet{StaffReportUS21}, a meme stock is characterized by a confluence of all these factors: large price moves, large volume changes, large short interest, frequent mentions on social media and significant coverage in the mainstream media. At the contrary, AAPL and MSFT are two non-meme stock to use as controls.\\

\subsection{Market data}
We download the time series with daily resolution of price and traded number of share (volume) from October 1$^{st}$ 2020 to June 30$^{th}$ 2021. For each stock we compute the daily returns:
\begin{equation}
    R_{t,s} = \frac{P_{t,s} - P_{t-1,s}}{P_{t-1,s}}
\end{equation}
where $P_{t,s}$ is the closing price of stock $s$ on day $t$. The daily trading volume is denoted as $Vol_{t,s}$.
\subsection{Reddit data}
\begin{figure}[ht]
        \centering
        \begin{subfigure}[b]{0.47\textwidth}
            \centering
            \includegraphics[width=\textwidth]{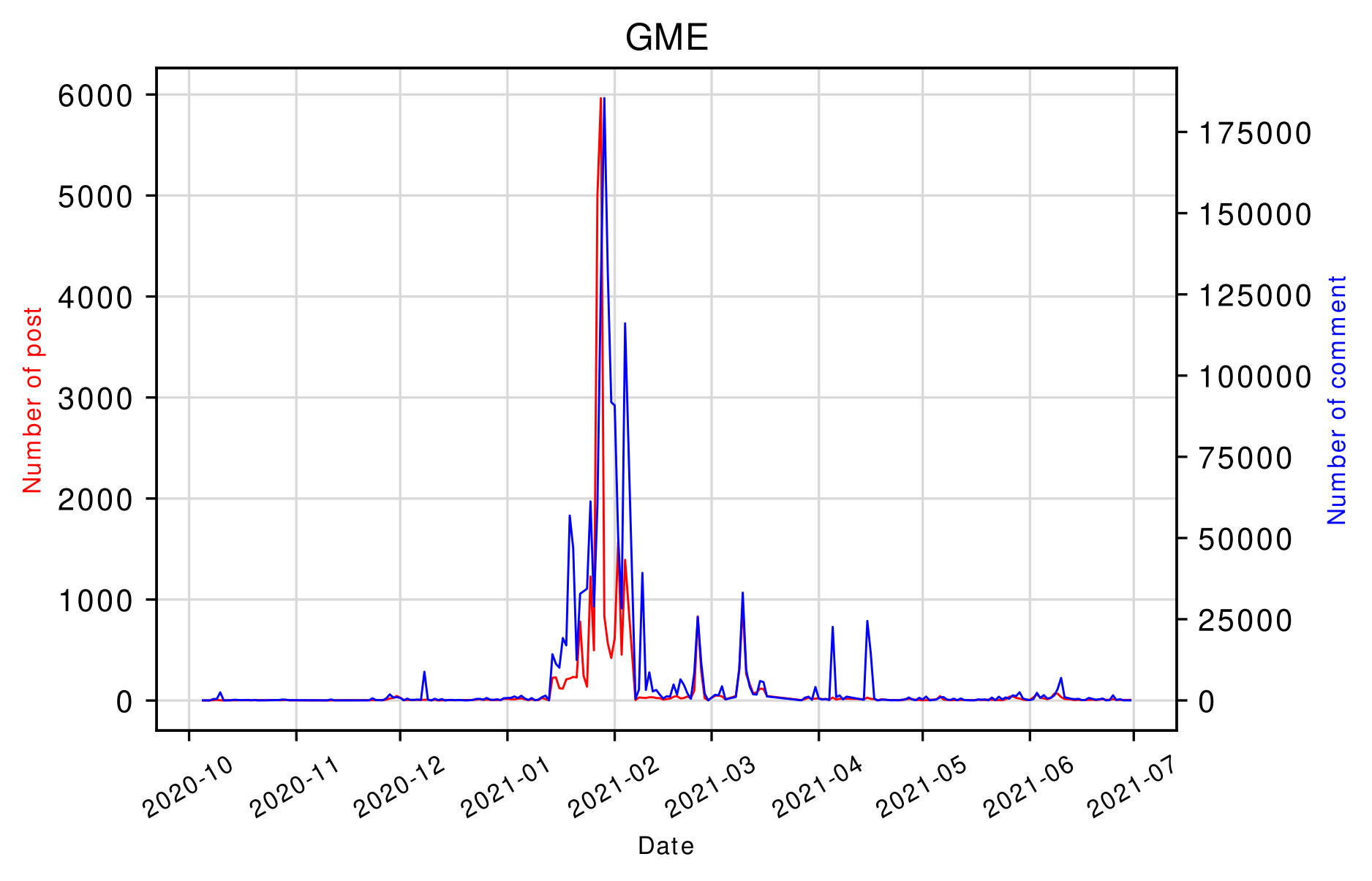}
        \end{subfigure}
        \hfill
        \begin{subfigure}[b]{0.47\textwidth}  
            \centering 
            \includegraphics[width=\textwidth]{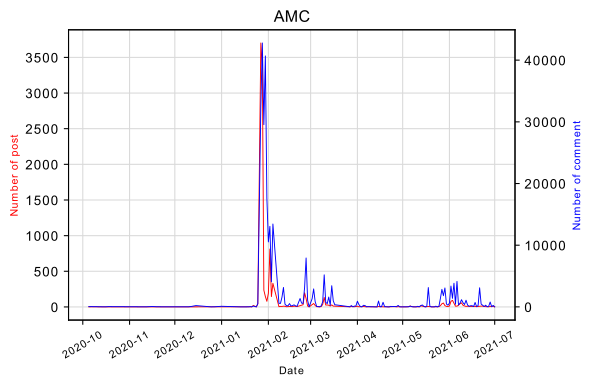}
        \end{subfigure}
        \begin{subfigure}[b]{0.47\textwidth}   
            \centering 
            \includegraphics[width=\textwidth]{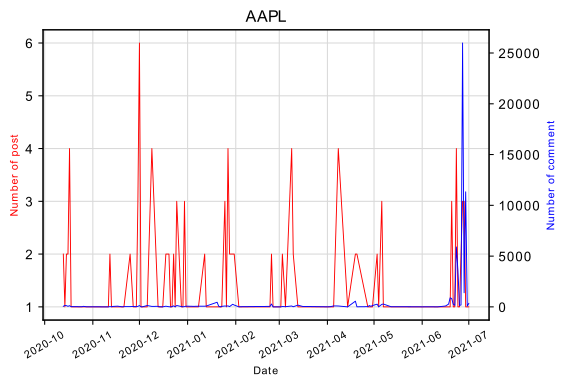}
        \end{subfigure}
        \hfill
        \begin{subfigure}[b]{0.47\textwidth}   
            \centering 
            \includegraphics[width=\textwidth]{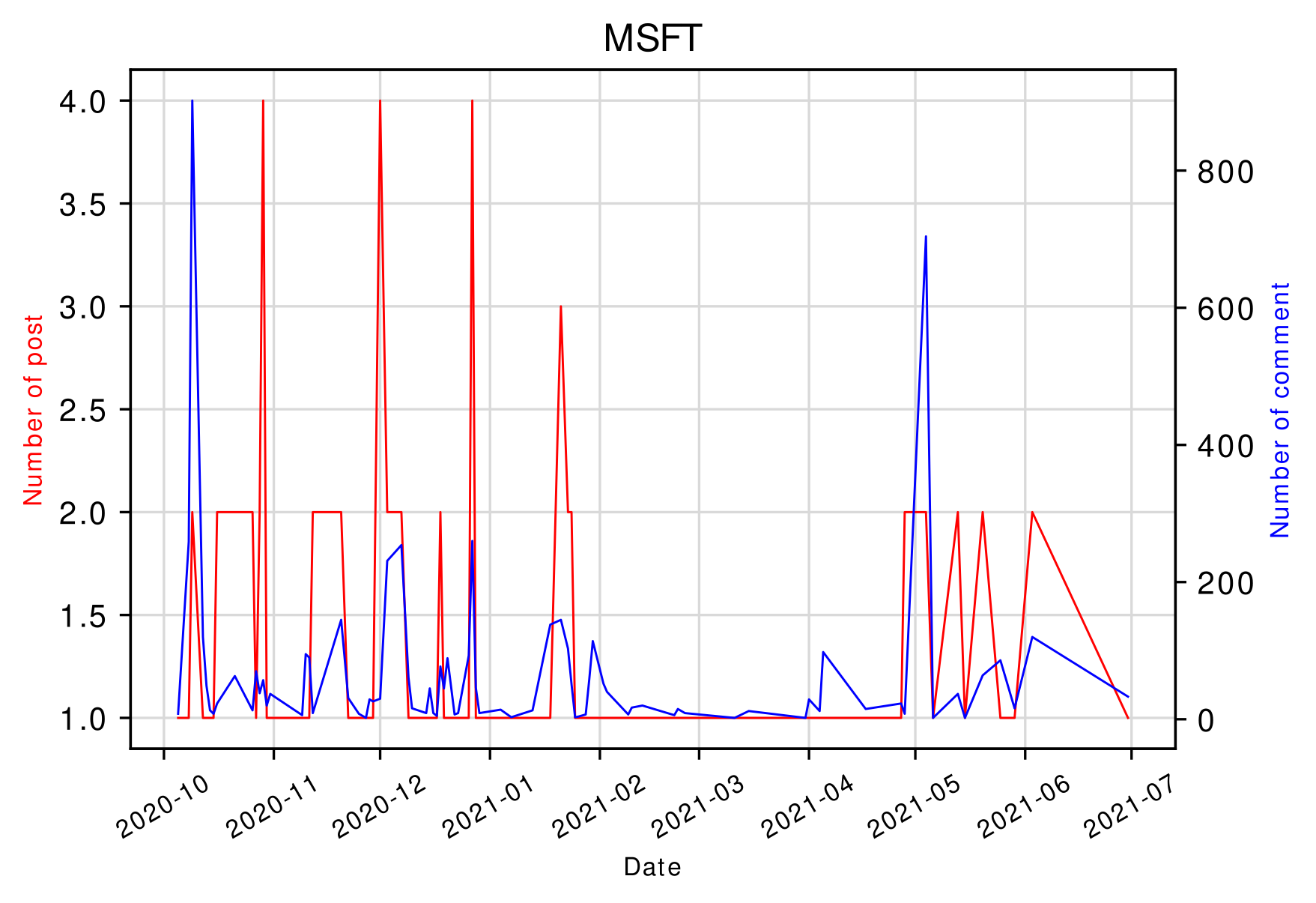}
        \end{subfigure}
        \caption[Time series of the daily number of post (submission) containing the word reported in the title of each subgraph (GME, AMC, AAPL, MSFT) and related comments on the subreddit \textit{r/Wallstreetbets} in the period October 2020 - June 2021. The post y-axis scale measure is on the left in red, while the corresponding scale for the comments is on the right of each graph and in blue.]
        {\small Time series of the daily number of post (submission) containing the word reported in the title of each subgraph (GME, AMC, AAPL, MSFT) and related comments on the subreddit \textit{r/Wallstreetbets} in the period October 2020 - June 2021. The post y-axis scale measure is on the left in red, while the corresponding scale for the comments is on the right of each graph and in blue.} 
        \label{post_comment}
    \end{figure}

\begin{figure}[ht]
        \centering
        \begin{subfigure}[b]{0.47\textwidth}
            \centering
            \includegraphics[width=\textwidth]{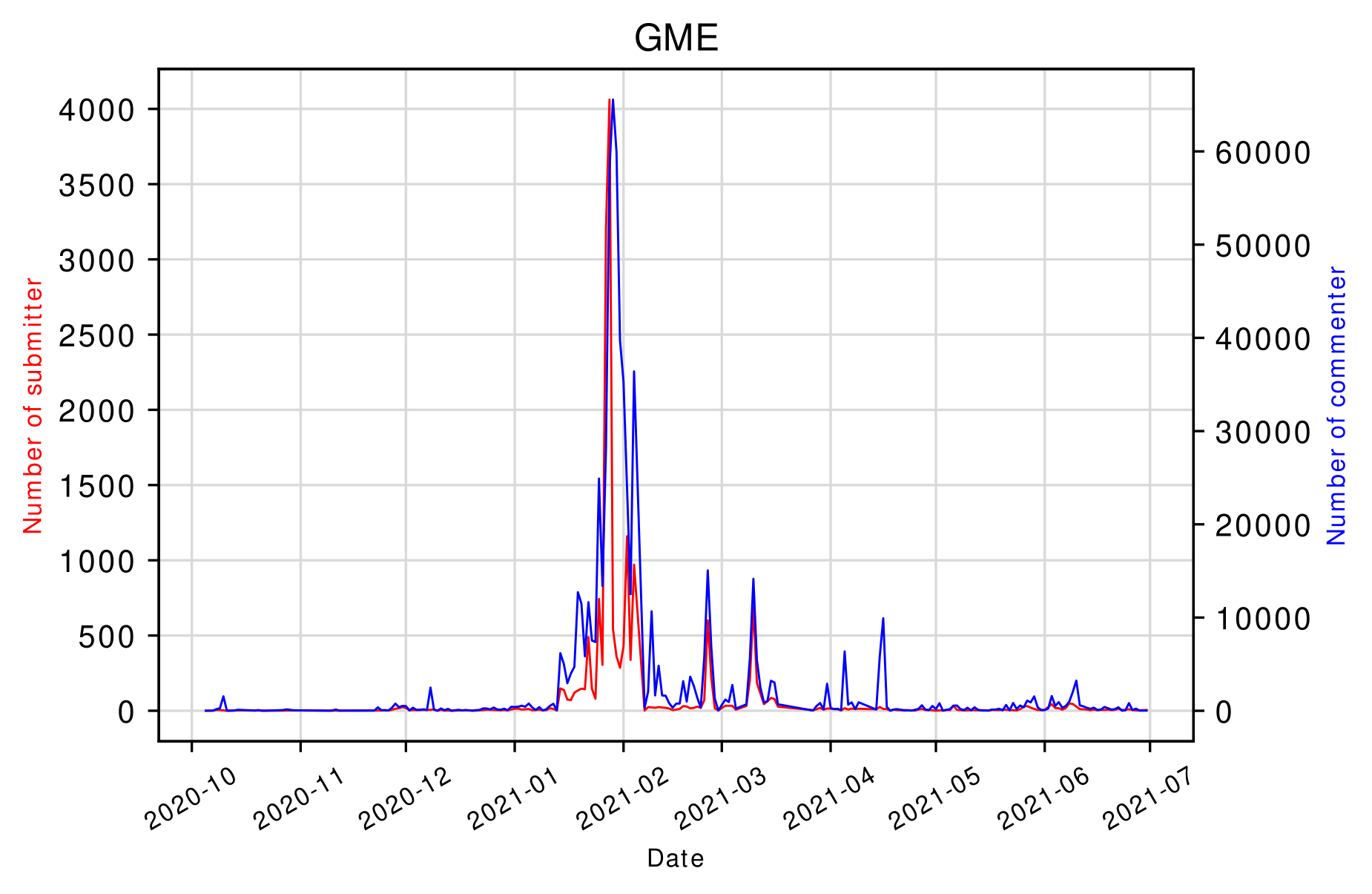}
        \end{subfigure}
        \hfill
        \begin{subfigure}[b]{0.47\textwidth}  
            \centering 
            \includegraphics[width=\textwidth]{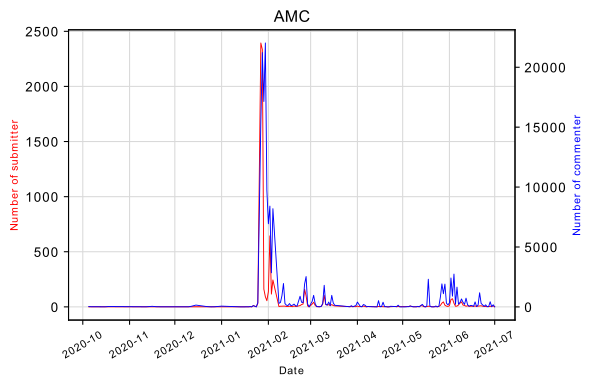}
        \end{subfigure}
        \begin{subfigure}[b]{0.47\textwidth}   
            \centering 
            \includegraphics[width=\textwidth]{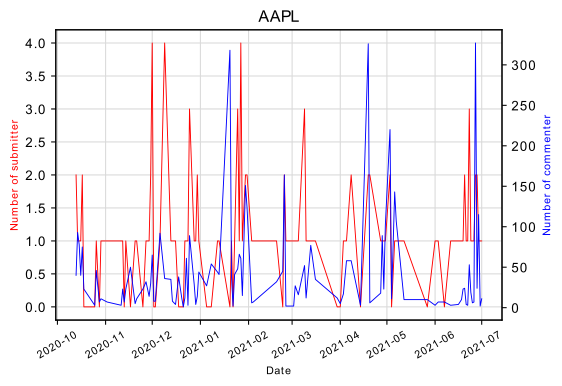}
        \end{subfigure}
        \hfill
        \begin{subfigure}[b]{0.47\textwidth}   
            \centering 
            \includegraphics[width=\textwidth]{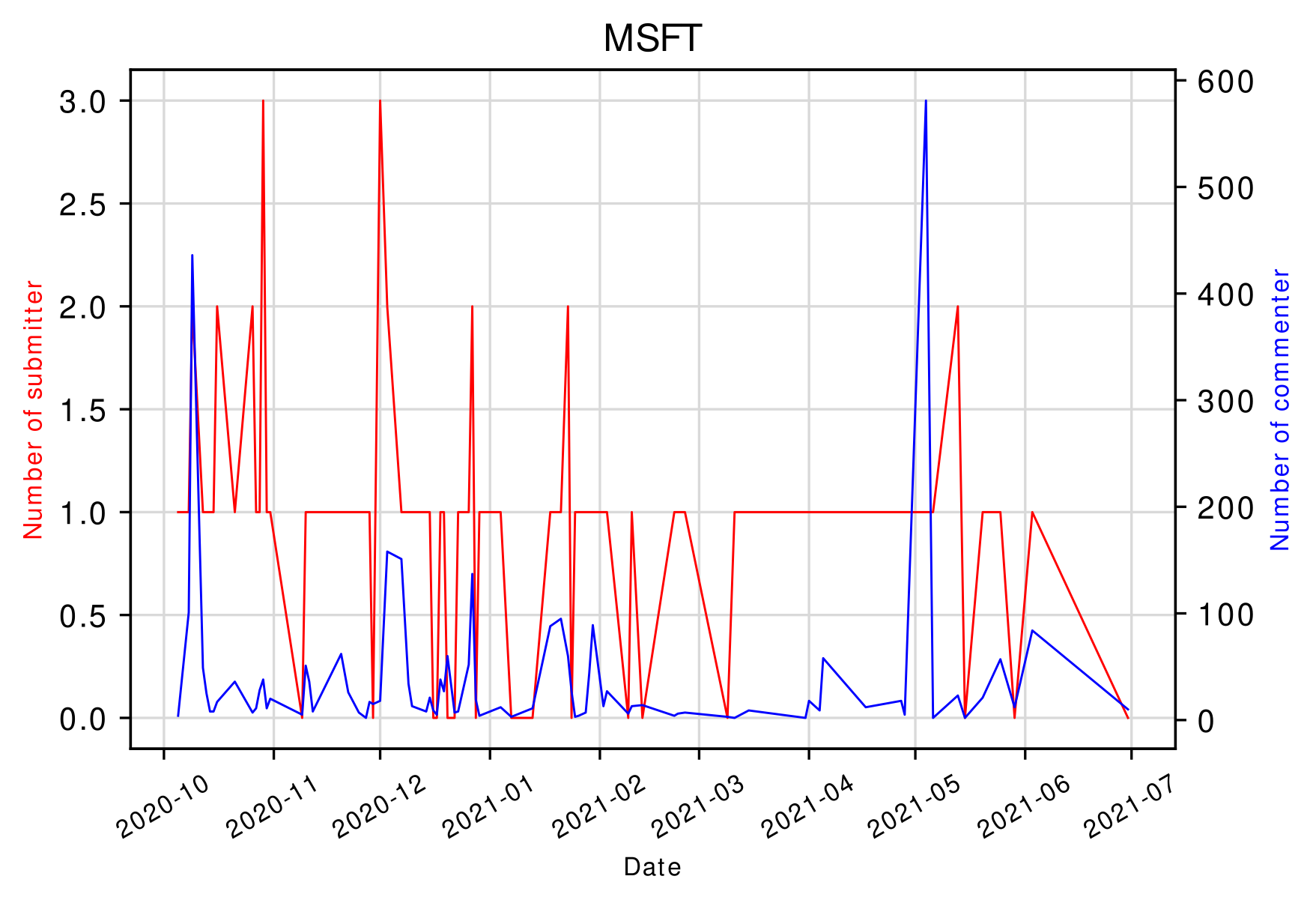}
        \end{subfigure}
        \caption[Time series of the daily number of submitters citing the word reported in the title of each subgraph (GME, AMC, AAPL, MSFT) and related comments by the users on the subreddit \textit{r/Wallstreetbets} in the period October 2020 - June 2021. The submitter y-axis scale measure is on the left in red, while the corresponding scale for the commenters is on the right of each graph and in blue.]
        {\small Time series of the daily number of submitters citing the word reported in the title of each subgraph (GME, AMC, AAPL, MSFT) and related comments by the users on the subreddit \textit{r/Wallstreetbets} in the period October 2020 - June 2021. The submitter y-axis scale measure is on the left in red, while the corresponding scale for the commenters is on the right of each graph and in blue.} 
        \label{submitter_commenter}
    \end{figure}

Rooting our analysis in the theoretical framework proposed by \cite{Pedersen2021}, we design a model to study the interaction of agents on the social platform, evaluate their coordination effort and quantify the impact of their action on the financial market. 
Modern social media contain an outstanding informative potential related to the users' sentiment evolution and opinion formation. If properly squeezed, the confidential information can be a forerunner for upcoming events.\\ Relying on the informative power of the social network and with the help of PRAW (an acronym for 'Python Reddit API Wrapper', a Python package that allows accessing Reddit's API), we downloaded all the posts containing as keyword the ticker of the stock to investigate. We limit the data download to the subreddit \textit{r/WallStreetBets} in the period from October 2020 to June 2021. For each post we obtained the related comment forests and attributes. We downloaded the data and run the analysis for the following keywords, standing for the stock tickers:  GME, AMC, AAPL, and MSFT. Precisely, we extracted every post (in Reddit jargon 'Submission') satisfying the conditions above and the related comment tree.\footnote{See the Appendix \ref{Appendix_data_download} for more details of downloaded data.}
We are interested in the emerging collective phenomena when the retail investors  cooperate to determine a significant effect on the financial market.\\
We exclude from our analysis the messages generated automatically from the bots (that in our dataset are denoted with the tag 'moderator' in the variable \texttt{distinguished}).
A conversation thread can be modeled as a directed tree $T_{t,s}^k = (M_{t,s}^k,C_{t,s}^k)$. $M_{t,s}^k$ the set of nodes represented by the messages in the tree $k$ (where the initial submission represents the root of the tree and the comments are the following-up branching) and $C_{t,s}^k$ is the set of edges, each of them connecting two messages linked by commenting activity. The direction of the edge points to the parent node to which the comment is addressed.\\
Figure \ref{post_comment} reports the number of daily posts (i.e. the number of trees) containing as keyword the ticker in the title of each subgraph and related comments (i.e. the number of nodes excluding the initial submission). Each subgraph has a double scale y-axis; on the left in red is the scale measuring the absolute value of daily submissions, on the right in blue is the corresponding measure for the number of related comments. We note that for the meme stocks (GME and AMC), the activity on the social network is massive compared to other well-known but not meme stocks (like AAPL and MSFT). Furthermore, when the social network is active on a specific topic, the correlation between creating new content and commenting activity is high.
Similarly, Figure \ref{submitter_commenter} shows the time series of the daily number of users who write content containing the ticker in the title of each subgraph and the daily number of users who take part in the conversation threads. In red on the left is the y-axis for the submitters; in blue on the right side is the corresponding one for commenters. The analysis is ideally in line with the one done for Figure \ref{post_comment}.\\
The social network data informativeness is not limited to its extent over time. It can be further squeezed by analyzing the interaction among users to identify their roles within the network. 
\subsubsection{Social network analysis}
\begin{figure}[ht]
     \centering
     \includegraphics[width=\textwidth]{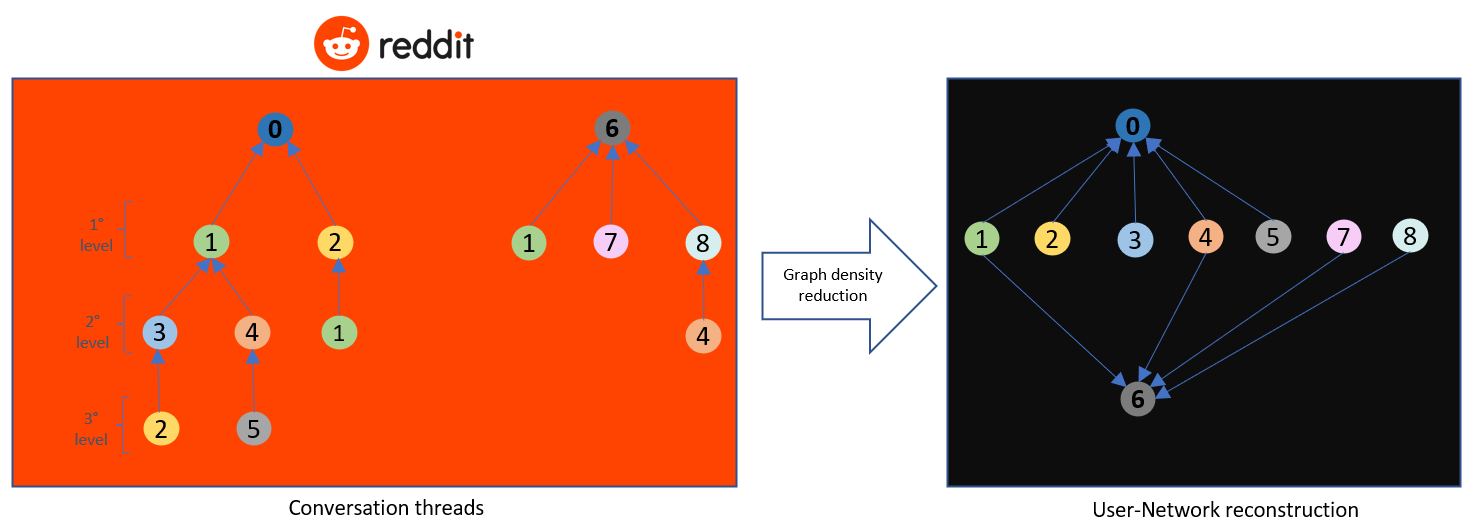}
     \caption{Example of user network reconstruction from two conversation threads. Starting from two conversation threads $T_{t,s}^{k} = (M_{t,s}^{k},C_{t,s}^{k}), k=1,2$ stylized as a comment trees (left-hand panel), we reconstruct the interaction among users $G_{t,s} = (N_{t,s},E_{t,s})$ by reducing the density and the level of detail. We consider all the comments to be addressed to the submission creators (in the example the users labeled with 0 and 6), as all the comments were at the first level of the tree. We do not consider how many times a user interacts with another one: we consider whether a user comments to the submitter to streamline the social network.
     }
     \label{network_rec}
 \end{figure}

\begin{figure}[ht]
    \centering
    \includegraphics[width=0.95\textwidth]{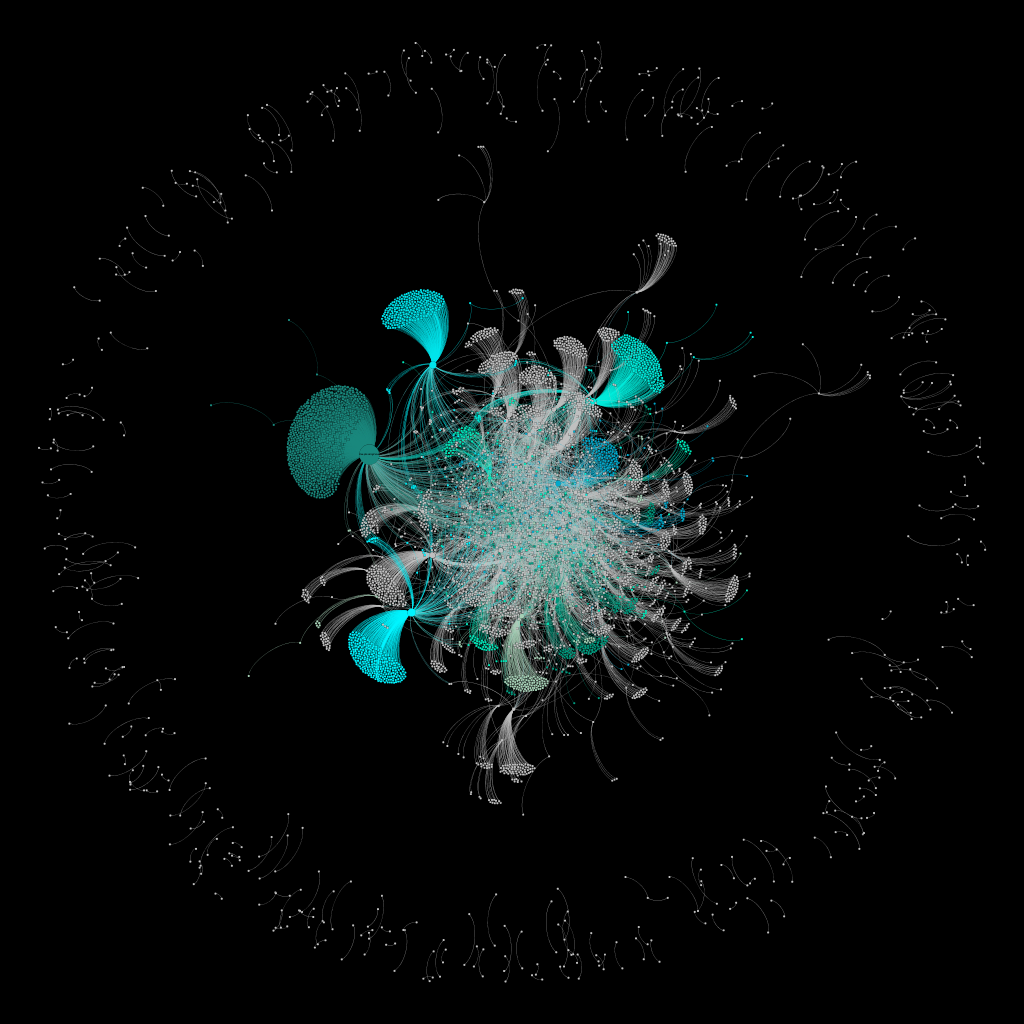}
    \caption{The Figure shows the network of users interacting on Reddit on January 14st, 2021. The network reports the interactions of users posting a submission containing the wording 'GME'. The network has 6,465 nodes and 8,741 edges. The top 10 nodes with the highest in-degree centrality are colored in blue.}
    \label{fig:Network_graph_GME_14_01_21}
\end{figure}

We extrapolate the daily agent interplay from the tree-level raw data by reconstructing their discussing interchanges. We define a network where the nodes represent the active users on the day $t$, and the edges are users' interactions through the comments: the network structure is outlined by $G_{t,s} = (N_{t,s},E_{t,s})$. The set of active users on the day $t$ for ticker $s$, $N_{t,s}$, represents the graph's nodes. The commenting activity establishes the links among them: $E_{t,s}$ is the set of directed edges from the user writing the comment to the author of the initial submission.\\
Following \cite{RiosEtAl2019}, we adopt a simplification in the user-network reconstruction (see Figure \ref{network_rec}): the links always point to the author of the initial submission even if the comment is a reply to a second or higher-order comment. The ratio underlying this network reduction is due to the need to identify the users acting like hubs, gaining popularity, consensus, and driving a potentially impactful movement with their contents. In addition, the stylization is still a realistic representation of the social dynamic: when scrolling the blog, the user first reads the main submission; then, if attracted by the topic, she opens the cascade of comments. Our reduction considers all the comments as first-level comments. This structure emphasizes the submitter that becomes the central actor of a thread and, eventually, the driver of a particular message or idea. 
Figure \ref{fig:Network_graph_GME_14_01_21} depicts an example of interaction among users. The network represents the interaction on Reddit on January 14st, 2021, based on submissions containing the ticker GME. There are 6,465 users (represented by the nodes) interacting on the platform throughout commenting activity (the 8,741 edges connecting them). The social network has a hub-and-spoke structure, with the colored users representing the hubs (i.e., central nodes in the network). In Appendix \ref{Network_graph} we report similar examples of networks for AMC, AAPL and MSFT.
\section{Methodology}
This section presents the backbone of our analysis. We describe the design and the functioning of the alert system to detect situations of misconducting behavior at social network level. Subsequently, we illustrate how we structure an event study analysis to check whether the alert system is capable of anticipating potential attempts of market manipulation.\\
\subsection{Alert system}
We devise an alert system based on social-network-retrieved information. Cooperation among users can translate into a dangerous impact on financial market stability. Detecting potential misconduct behaviors and anticipating a coordinated action concocted on social media might be beneficial for the financial market well-functioning.
The alert system is query-dependent, meaning that we have to instruct the system on which keyword we want to monitor. For the sake of practical usability, two consecutive stpng compose the structure. \\
\subsubsection{First stage}
The first stage of the alert system consists of a screening of the days where the ticker-related activity on the social network is extensive compared to the previous days. We use the volume-related metrics to implement the first stage: we skim the days, identifying when ticker-related activity is extraordinary.
Technically speaking, for every day and each ticker, we determine:
\begin{itemize}
\item The number of submissions citing the ticker in their content;
\item The number of active users discussing that ticker;
\item The overall\footnote{Meaning the whole activity during the day on the subreddit \textit{r/WallStreetBets}, not specifically related to a ticker.} activity on the subreddit, both in terms of submissions and users. 
\end{itemize}
And construct the following variables:
\begin{enumerate}
\item[(1)] Relative number of daily submission: the ratio between the submissions citing ticker and the total submissions on the subreddit, to identify the portion of the activity on that topic during the day;
\item[(2)] Absolute number of daily submission: the total number of daily submissions about that ticker, to identify the magnitude of the movement in absolute terms;
\item[(3)] Relative percentage change in the number of daily submissions:  the percentage variation of number of ticker-related submissions with respect to the previous day;
\item[(4)] Relative number of daily users: the ratio between the number of users citing ticker and total users on the subreddit, to identify the portion of the community discussing that topic during the day;
\item[(5)] Absolute number of daily users: computed as the total number of daily users citing that ticker, to identify the magnitude of the user-trend in absolute terms;
\item[(6)] Relative percentage change in the number of daily users:  the percentage variation of number of ticker-related users with respect to the previous day;
\end{enumerate}
The first stage of the alert system switches on when at least four variables exceed their threshold. 
For the relative and the absolute number of daily submissions (users), variables (1), (2), (4), (5), the threshold for the day $t$ is the mean value of the variable over the previous ten days plus one mean absolute deviation computed over the same period.
We define the mean value of the variable as
\begin{equation}
    \bar{V}_{t,s}^{(v)} = \frac{1}{I}\sum_{i=1}^{I} V_{t-i,s}^{(v)}
\end{equation}
The mean absolute deviation as
\begin{equation}
    MAD({V}_{t,s}^{(v)}) = \frac{1}{I}\sum_{i=1}^{I} \left( V_{t-i,s}^{(v)} - \bar{V}_{t,s}^{(v)} \right)
\end{equation}
Hence, the threshold is defined as 
\begin{equation}\label{threshold}
    T_{t,s}^{(v)} = \bar{V}_{t,s}^{(v)} + MAD({V}_{t,s}^{(v)})
\end{equation}
where $v$ refers to the variables $j = 1,2,4,5$, $t$ indicates the day, $s$ the ticker $s=GME,AMC,AAPL,MSFT$ and $I$ is the length of the window over which we compute the threshold, in our case $I=10$.
For the relative percentage change in the number of daily submissions (users), variables (3) and (6), the threshold to overcome is $100\%$, hence when on the day $t$ they double the value with respect to the previous day $t-1$.\\
When the social network conditions determine the activation of the alert, we approach the situation in a prudential and conservative way: we keep monitoring the stock until the average number of active indicators over the previous three days is below three. In this way, we keep controlling the situation even if it is not exceptional compared to the earlier days, but it is still turbulent.\\
A single indicator (or variable) switches on when it overcomes its critical threshold defined in (\ref{threshold}):
\begin{equation}
    I_{V_{t,s}^{(v)}} = 
    \begin{cases}
    1, & \mbox{if } V_{t,s}^{(v)} > T_{t,s}^{(v)}\\
    0, & \mbox{otherwise}
    \end{cases}
\end{equation}
The minimum alert-activation condition is:
\begin{equation}
    \sum_{v=1}^6  I_{V_{t,s}^{(v)}} \ge 4
\end{equation}
The alert remains on until the following condition verifies:
\begin{equation}
    \frac{1}{3} \sum_{i=1}^3 \sum_{v=1}^6 I_{V_{t-i,s}^{(v)}} \le 3
\end{equation}
Step one of the alert system detects the days when the activity is exceptional, calling for further controls on the network structure. 
\subsubsection{Second stage}
The second stage of the alert system activates only for those days recognized as an alert state by the first one.\\
For each day selected by the first step, we reconstruct the network structure to model the interaction among the agents, $G_{t,s} = (N_{t,s},E_{t,s})$ in the manners set in the previous paragraph: the links always point to the author of the initial submission even if the comment is a reply to a second or higher-order comment.\\
We also implement other filters in the network modeling: creating a nowcasting alert system requires the tool to be quick and smartly devised, beyond the fact that it has to detect the bigwigs. Hence, the network reconstruction is made of only the users whose submission obtained a score above the median and with a cascade of at least ten comments. \\
We now move to the detection of the users acting like leaders, able to gain trustworthiness, popularity, prestige, leadership, and authority, essential features for a  virtual user to lead a movement that can translate its effects on the real economy.\\
We rank the users according to their in-degree centrality (fraction of nodes its incoming edges are connected to) for each day in the subset of first-stage-detected days:
\begin{equation}
    C_D(n_{t,s}^{(i)}) = \frac{\sum_j a_{t,s}^{ij}}{N_{t,s}}
\end{equation}
The indicator identifies the ten authors with the highest relative incoming links: users able to attract a vast portion of the community. To test whether the detected authors are trending and critically acclaimed, for each day $t$, we define a set of the most critical users according to the algorithm of PageRank \citep{PageRank99}. We consider the network in the window $[t-20,t-1]$ and identify the users who were standing out for the published content. According to the ranking, the first twenty users belong to the set of influencers at time $t$. We finally check whether some of the top in-degree centrality authors on the day $t$ belong to the influencers set over the past twenty days\footnote{We do some robustness checks by changing the number of users with the highest in-degree centrality, and also modifying the parameters of the influencer set (window length and number of critical users identified by the PageRank algorithm). The results are unaffected.  }. If the intersection between the two sets is not empty, the second stage of the alert system switches on.\\
The methodology narrows down the set of agents we monitor to avoid misconducting behaviors on the online social network and prevent repercussions on the financial stability. The method aims at pin down the users who might manage suspicious movements by promoting extreme investment strategies masked by financial pieces of advice. In a perspective of macroprudential stability, a regulatory authority using the tool can quickly pinpoint the users to check to analyze their contents throughout textual analysis tools and eventually ban the profiles.

\subsection{Analysis of abnormal returns}
We finally check whether the algorithm based on social network analysis can adequately detect the financially unstable days. 
In order to evaluate the accuracy of our algorithm, we develop an event study analysis following \citet{mackinlay1997event}. We measure whether abnormal returns occur for the stock discussed in the Reddit community right after the alert turns on. \\
The abnormal returns are constructed by defining an estimation window that goes from $T_0$ to $T_1$, and an event date $\tau$. The event window goes from $T_1$ to $T_2 =$. $L_1 = T_1 - T_0$ and $L_2 = T_2 - T_1$ are respectively the length of the estimation and the event window. The abnormal returns are defined with the specification of the market model, to purge it by the market fluctuations:
\begin{equation}\label{eq:mkt_mod}
A R_{\tau,s}=R_{\tau,s}-\hat{\alpha}_{s}-\hat{\beta}_{s} R_{\tau,m}
\end{equation}
where $R_{\tau,s}$ is the return for security $s$ at time $\tau$, while $R_{\tau,m}$ is the market return. The parameters of the model are estimated via OLS over the estimation window $[T_0:T_1]$.
Under the null hypothesis, the abnormal returns, $AR_{\tau,s}$, are normally distributed with zero conditional mean and conditional variance $\sigma^{2}\left(A R_{ \tau,s}\right)$ is set to be the variance of the OLS residuals $\sigma_{\varepsilon_{t}}^{2}$.


Under the null hypothesis, that event has no impact on the returns, the distribution of the abnormal returns in the event window is:
\begin{equation}\label{eq:ar_distr}
A R_{\tau,s} \sim N\left(0, \sigma^{2}\left(A R_{\tau,s}\right)\right)
\end{equation}
The significance of the abnormal returns is tested via the
non-parametric rank test by \citet{corrado1989nonparametric}. The abnormal returns are standardized as a ranked descending variable defined by:
\begin{equation}
K_{\tau,s}=\frac{\operatorname{rank}\left(A R_{\tau,s}\right)}{1+M_s}
\end{equation}
where $M_s$ are the the number of observations in the sample for security $s$. The ranked variable is uniformly distributed in a [0,1] interval.\\
The variance computed across all the stock $s$ observations is:
\begin{equation}
S_{\bar{K}}^{2}=\frac{1}{M_s} \sum_{t=0}^{T} \left({K}_{s,t}-0.5\right)^{2}
\end{equation}
where 0 and T stand for the first and last date of the sample. The significance is computed as a t-rank statistic for a standard uniform distribution with expected value of 0.5:
\begin{equation}
t_{\text {rank }}=\left(\frac{{K}_{s, \tau}-0.5}{S_{{K}}}\right)
\end{equation}
These results can be used to make inference on the absolute returns for every security in the event window.\\ 
We also test the impact of the detected events on the trading volume. Specifically, we consider the abnormal trading volume computed as the volume traded on a given day divided by the average volume traded on the estimation window [$T_0:T_1$]. For each day $\tau$, the abnormal trading volume is defined as:
\begin{equation}
    AVol_{\tau,s} = \frac{Vol_{\tau,s}}{\frac{1}{L_1}\sum_{t=T_0}^{T_1}Vol_{t,s}}
\end{equation}
\section{Results}
This section consists of an empirical application of the methodology we design in the previous section. We present the days spotted by the alert tool and evaluate the associated abnormal returns with an event study analysis. Subsequently, we provide a regression analysis to understand the main drivers of abnormal returns and how they differ between the meme and non-meme stocks. We find that social networks-related variables significantly explain the meme stock performance, while market-related variables primarily drive non-meme stocks.
\subsection{Event detection}
\begin{figure}[ht]
    \centering
    \includegraphics[width=\textwidth]{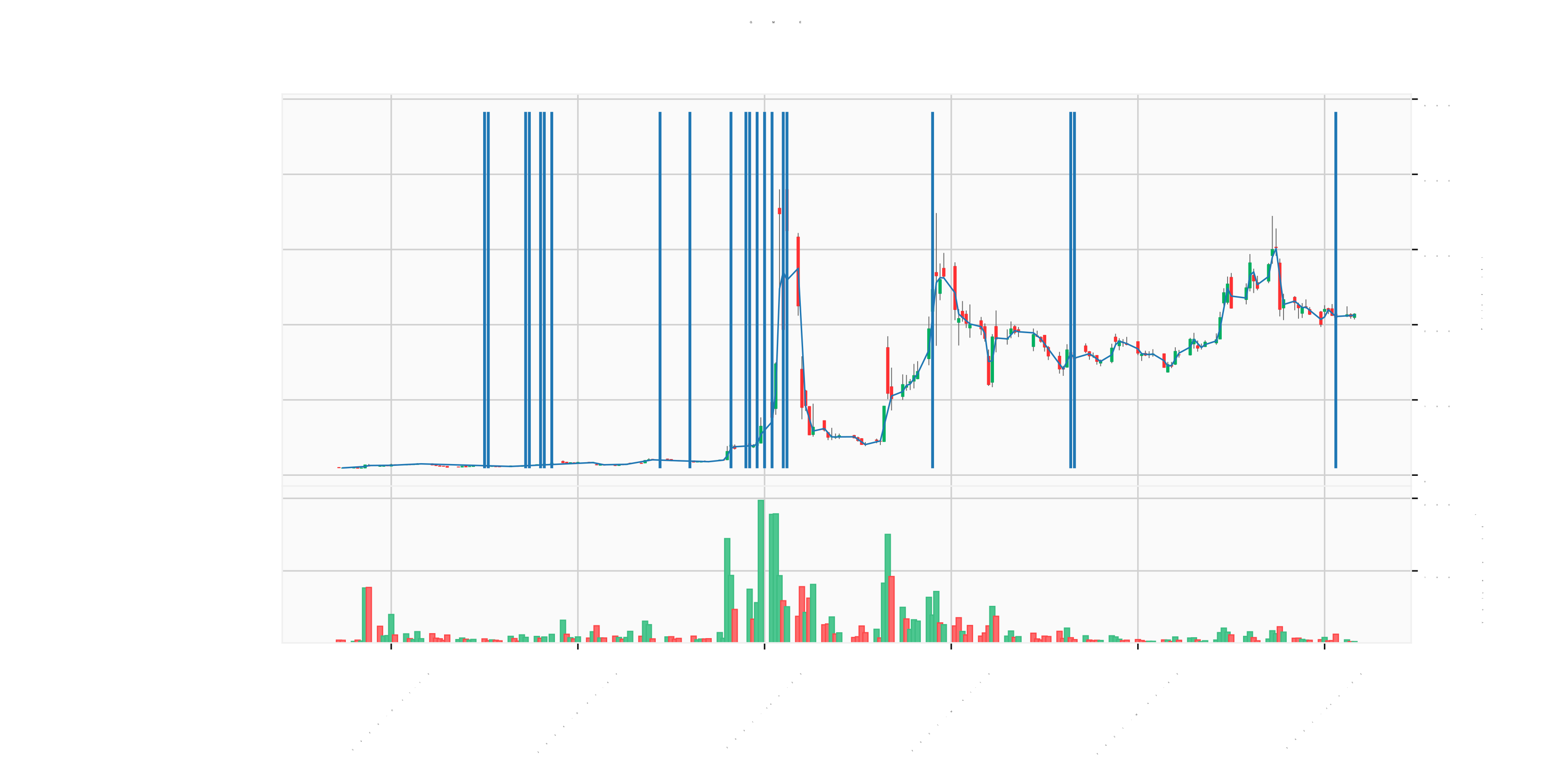}
    \caption{The figure shows financial data together with the alert days for the stock GME over the period October 2020 - June 2021. The above panel of the figure presents the time series of the daily price with a candlestick chart and the vertical blue lines are the days when the alert system turns on; in the bottom panel, the daily traded volume over the corresponding period.}
    \label{fig:PriceVolAlert_GME}
\end{figure}

\begin{figure}[ht]
    \centering
    \includegraphics[width=\textwidth]{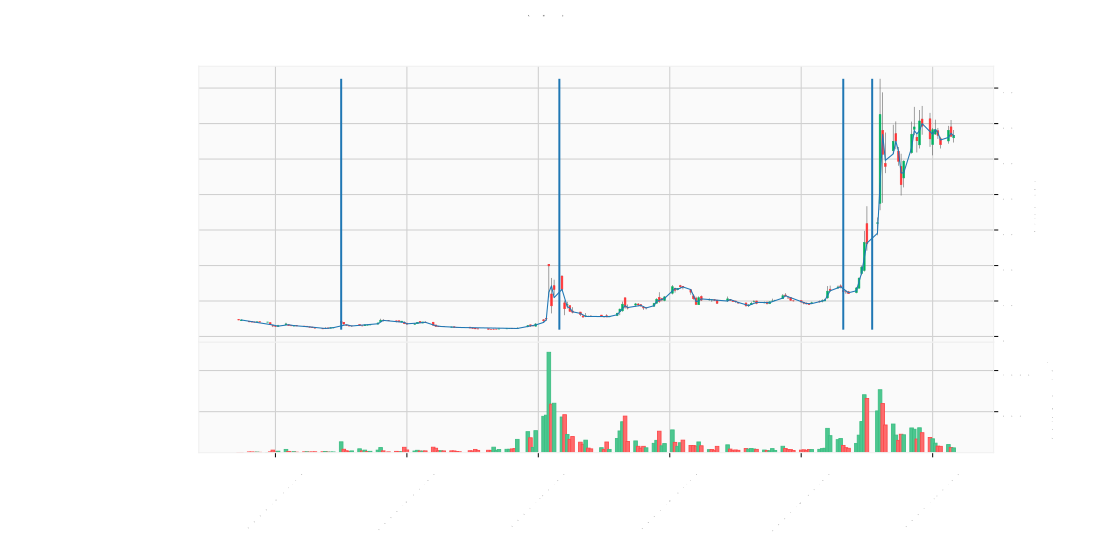}
    \caption{The figure shows financial data together with the alert days for the stock AMC over the period October 2020 - June 2021. The above panel of the figure presents the time series of the daily price with a candlestick chart and the vertical blue lines are the days when the alert system turns on; in the bottom panel, the daily traded volume over the corresponding period.}
    \label{fig:PriceVolAlert_AMC}
\end{figure}

\begin{figure}[ht]
    \centering
    \includegraphics[width=\textwidth]{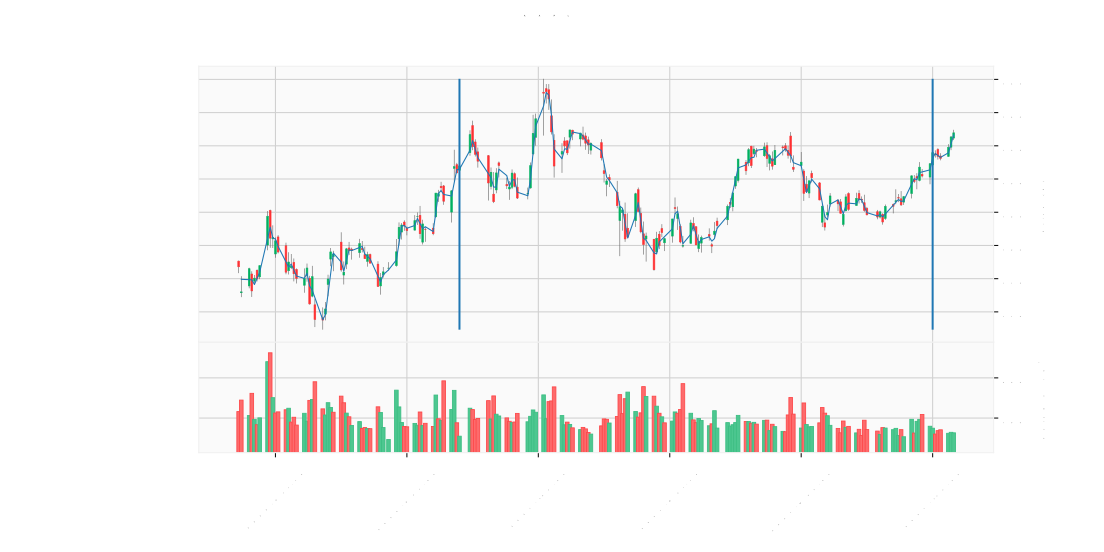}
    \caption{The figure shows financial data together with the alert days for the stock AAPL over the period October 2020 - June 2021. The above panel of the figure presents the time series of the daily price with a candlestick chart and the vertical blue lines are the days when the alert system turns on; in the bottom panel, the daily traded volume over the corresponding period.}
    \label{fig:PriceVolAlert_AAPL}
\end{figure}

\begin{figure}[ht]
    \centering
    \includegraphics[width=\textwidth]{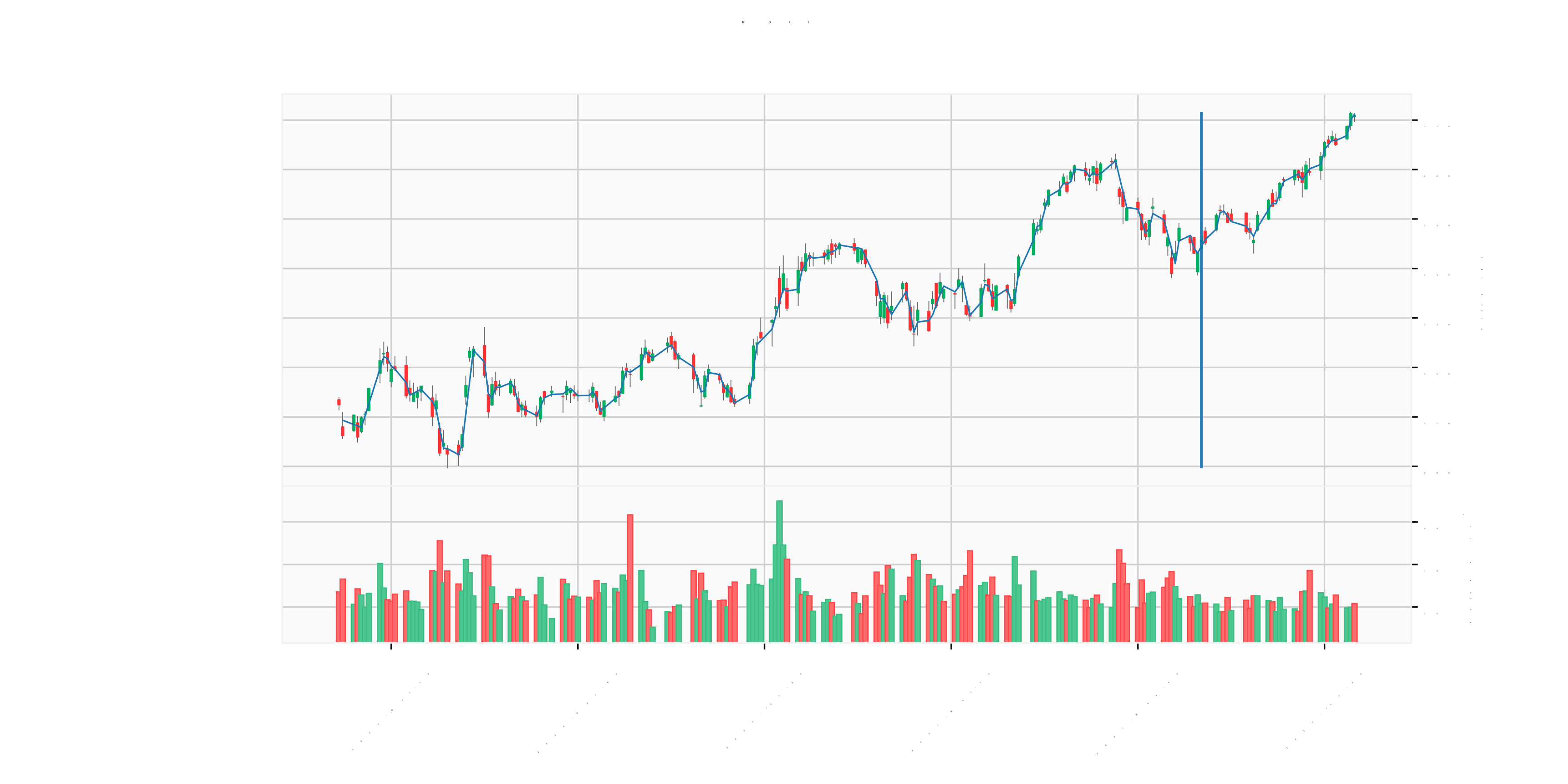}
    \caption{The figure shows financial data together with the alert days for the stock MSFT over the period October 2020 - June 2021. The above panel of the figure presents the time series of the daily price with a candlestick chart and the vertical blue lines are the days when the alert system turns on; in the bottom panel, the daily traded volume over the corresponding period.}
    \label{fig:PriceVolAlert_MSFT}
\end{figure}

In Figures \ref{fig:PriceVolAlert_GME}, \ref{fig:PriceVolAlert_AMC}, \ref{fig:PriceVolAlert_AAPL}, and \ref{fig:PriceVolAlert_MSFT} we report the daily time series of price and traded volume for the stocks under analysis, together with the the days the alert system detects extraordinary activity and cooperation on the social network.\\ 
The alert tool identifies 21 suspicious movements on the social network for GME, 4 for AMC, 2 for AAPL, and 1 for MSFT.\\
GME is the stock facing the most incredible attention and chattering on social media. Let us consider the short squeeze that happened at the end of January 2020 as the most striking episode of social network cooperation impacting the financial economy. We appreciate that our alert tool can detect abnormal and suspecting movements on the network many days before the social coordination affects the financial market. In addition, the tool can narrow down the few users that drive the movement, simplifying the eventual inspection by a surveillance authority. In Table \ref{Tab:Event_date_user}, we report all the dates and the users noticed by the alert system. The tool can trace the user acting like leaders for the movement. In Appendix \ref{Network_graph}, we show a user network to highlight how agents interacts on the social network.\\
Despite the lower media frenzy compared to GME, AMC experiences considerable attention on social media. Retail investors attempt a pump-and-dump scheme during the same days of the GME frenzy, but they create actual instability on the second mid of May 2021 when they considerably purge the price. The price has risen about seven times in ten days since the warning first lit on May 19th, 2021.\\
Unlike GME and AMC, the financial performance of prices and volumes of non-meme-stocks (AAPL and MSFT) is entirely unaffected by social media activity. Leading users ('fanatics') on social networks are aware that stocks with such high capitalization are challenging to implement pump-and-dump schemes or drive the price away from fundamental value. For this reason, the alert is rarely triggered and the identified alerting events are insignificant compared to those of meme-stocks.\\

\begin{table}[h!]
\centering
\footnotesize
\begin{tabular}{lllll}
\multicolumn{2}{c}{\textbf{GME}}                                                          &                      & \multicolumn{2}{c}{\textbf{AMC}}                                                              \\ \cline{1-2} \cline{4-5} 
Alert date ($\tau$) & Influencer(s)                                                       &                      & Alert date ($\tau$) & Influencer(s)                                                           \\ \cline{1-2} \cline{4-5} 
09/11/2020          & DeNovaCain                                                          &                      & 09/11/2020          & Killtrend                                                               \\
10/11/2020          & Veryforestgreen                                                     &                      & 31/01/2021          & \begin{tabular}[c]{@{}l@{}}dhiral1994\\ BrandinoGames\end{tabular}      \\
20/11/2020          & \begin{tabular}[c]{@{}l@{}}Neothedreamer\\ Imboredsoyh\end{tabular} &                      & 19/05/2021          & Realplayer16                                                            \\
21/11/2021          & Ackilles                                                            &                      & 01/06/2021          & Nobjos                                                                  \\
22/11/2021          & Ackilles                                                            &                      &                     &                                                                         \\
25//11/2021         & Ackilles                                                            &                      &                     &                                                                         \\
27/11/2020          & SIR\_JACK\_A\_LOT                                                   &                      &                     &                                                                         \\
26/12/2020          & Uberkikz11                                                          &                      &                     &                                                                         \\
03/01/2021          & Uberkikz11                                                          &                      &                     &                                                                         \\
14/01/2021          & DeepFuckingValue                                                    &                      &                     &                                                                         \\
18/01/2021          & Its-Loki                                                            &                      &                     &                                                                         \\
19/01/2021          & \begin{tabular}[c]{@{}l@{}}Gardeeon\\ DeepFuckingValue\end{tabular} &                      &                     &                                                                         \\
21/01/2021          & Unlucky-Prize                                                       &                      &                     &                                                                         \\
23/01/2021          & Unlucky-Prize                                                       &                      &                     &                                                                         \\
25/01/2021          & DeepFuckingValue                                                    &                      &                     &                                                                         \\
28/01/2021          & DeepFuckingValue                                                    &                      &                     &                                                                         \\
29/01/2021          & DeepFuckingValue                                                    &                      &                     &                                                                         \\
09/03/2021          & dumbledoreRothIRA                                                   &                      &                     &                                                                         \\
15/04/2021          & OPINION\_IS\_UNPOPULAR                                              &                      &                     &                                                                         \\
16/04/2021          & DeepFuckingValue                                                    &                      &                     &                                                                         \\
25/06/2021          & Chillznday                                                          &                      &                     &                                                                         \\ \cline{1-2} \cline{4-5} 
\multicolumn{2}{c}{\# events = 21}                                                        &                      & \multicolumn{2}{c}{\# events = 4}                                                             \\
                    &                                                                     &                      &                     &                                                                         \\
\multicolumn{2}{c}{\textbf{AAPL}}                                                         &                      & \multicolumn{2}{c}{\textbf{MSFT}}                                                             \\ \cline{1-2} \cline{4-5} 
Alert date ($\tau$) & Influencer(s)                                                       &                      & Alert date ($\tau$) & Influencer(s)                                                           \\ \cline{1-2} \cline{4-5} 
24/12/2020          & Nafizzaki                                                           &                      & 20/05/2021          & \begin{tabular}[c]{@{}l@{}}Mysterious----\\ EmphasisOk3036\end{tabular} \\
22/06/2021          & Tilthefatladysings                                                  &                      &                     &                                                                         \\ \cline{1-2} \cline{4-5} 
\multicolumn{2}{c}{\# events = 2}                                                         & \multicolumn{1}{c}{} & \multicolumn{2}{c}{\# events = 1}                                                            
\end{tabular}
\caption{The Table reports for each stock under analysis the events spotted by the alert system. For each detected event, it indicates the date and the and the user(s) under investigation.}
\label{Tab:Event_date_user}
\end{table}
\subsection{Analysis of the abnormal returns}
The submissions from the influencers during the days of unusual activity, detected by the two stpng of the alert system, are considered as events potentially triggering a response in the financial market. 
We analyze abnormal returns constructed with a market model to evaluate whether the submission from the influencers triggers a reaction in the financial markets.
For each detected event, i.e., when the alert system turns on ($\tau = 0$), we compute the abnormal return on the estimation window  [-21:-11], $L_1 = 10$ and we consider as event window the period [-10:10], $L_2 = 21$ trading days. The abnormal return is defined with a market model, specified in (\ref{eq:mkt_mod}), with the CRSP index as a proxy for the market returns. We impose a minimum of 10 days between two events to avoid contagion on the event window and consider the events independent. If an event happens on a non-trading day, we consider the next trading day as the event date.\\
Under these conditions, we end up with eight events for GME, four events for AMC, two events for AAPL, and one for MSFT over the sample period. It is clear that influencers driving unusual activity on social are more present for memes than non-meme stocks.\\
Considering that Reddit users are mostly inexperienced retail investors, we assume they primarily trade with a long position on the stock. Indeed, their primary concern is to move the price up by massively buying the stock. For this reason, we report and evaluate the significance of the event that generates the highest positive abnormal returns in the ten days after. 
Results are reported only for the case of meme stock. The event for the non-meme stock does not generate a clear upward trend after the submission, although some abnormal returns are significant. The events reported for GME and AMC are respectively submissions by \textit{u/DeepFuckingValue} on January 14th, 2021 and by \textit{u/realplayer16} on May 19th, 2021. 
\begin{table}[H]
\centering
\setlength{\tabcolsep}{10pt}
\begin{tabular}{cccc} \toprule
{\hspace{60pt}} &   \\ 
 Day ($\tau$) & $AR_{GME,t}$   &    $CAR_{GME,t}$  & $AVol_{GME,t}$ \\ 
\midrule

-10 & 0.012 & 0.012 &0.464 \\

-9  & -0.009 & 0.004 &0.651\\

-8 & -0.075 & -0.072 & 1.001\\

-7 & 0.013 & -0.057 &  0.494\\

-6 & $0.057^{}$ & -0.0009 &0.551 \\

-5  & -0.011 & -0.011 & 0.554\\

-4 & -0.043 & -0.054 & 0.492\\

-3 & $0.104^{}$ & 0.049 & 1.060\\

-2 & -0.038 &  0.011 &0.568\\

-1  & $0.534^{**}$ & 0.545 & 11.521 \\

0 & $0.233^{*}$ & 0.778  & 7.379\\

+1 & -0.152 & 0.626  & 3.801\\

+2 & $0.078^{}$ & 0.704 & 6.117\\

+3  & -0.030 & 0.674  & 2.657\\

+4 & $0.082^{}$ & 0.756  &4.571\\

+5 & $0.493^{*}$ & 1.249  &17.338\\

+6 & $0.164^{*}$ & 1.412  & 16.173\\

+7 & $0.935^{**}$ & 2.347 & \textbf{20.292}\\

+8  & \textbf{1.349}$^{**}$ & 3.697 &11.961 \\

+9 & -0.443$^{**}$ & 3.253 &7.463\\

+10 & $0.630^{**}$ & \textbf{3.883} & 2.503\\

\bottomrule

\end{tabular}
\caption{Event study for GME on the event date January 14th, 2021 ($\tau = 0$). This table shows the abnormal returns, the cumulative abnormal returns and the abnormal volumes on a [-10:+10] days event window around the event day. $^{***}$, $^{**}$, $^{*}$ represents significance at 1\%, 5\% and 10\% for the abnormal returns only. The highest value for each column in bold.}
\label{table:ar_gme}
\end{table}

\begin{figure}[ht]
        \centering
        \begin{subfigure}[b]{0.49\textwidth}
            \centering
            \includegraphics[width=\textwidth]{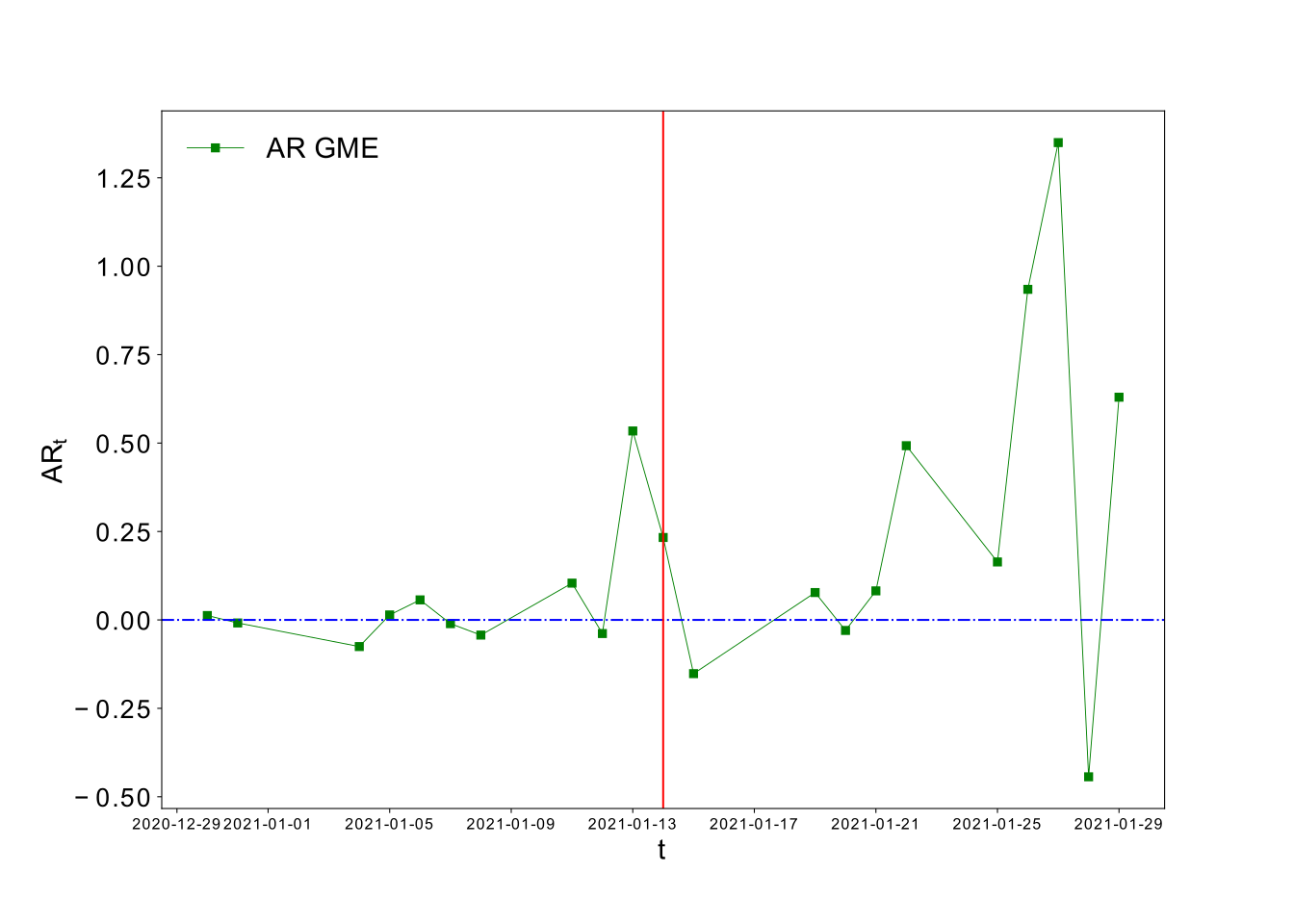}
        \end{subfigure}
        \hfill
        \begin{subfigure}[b]{0.49\textwidth}  
            \centering 
            \includegraphics[width=\textwidth]{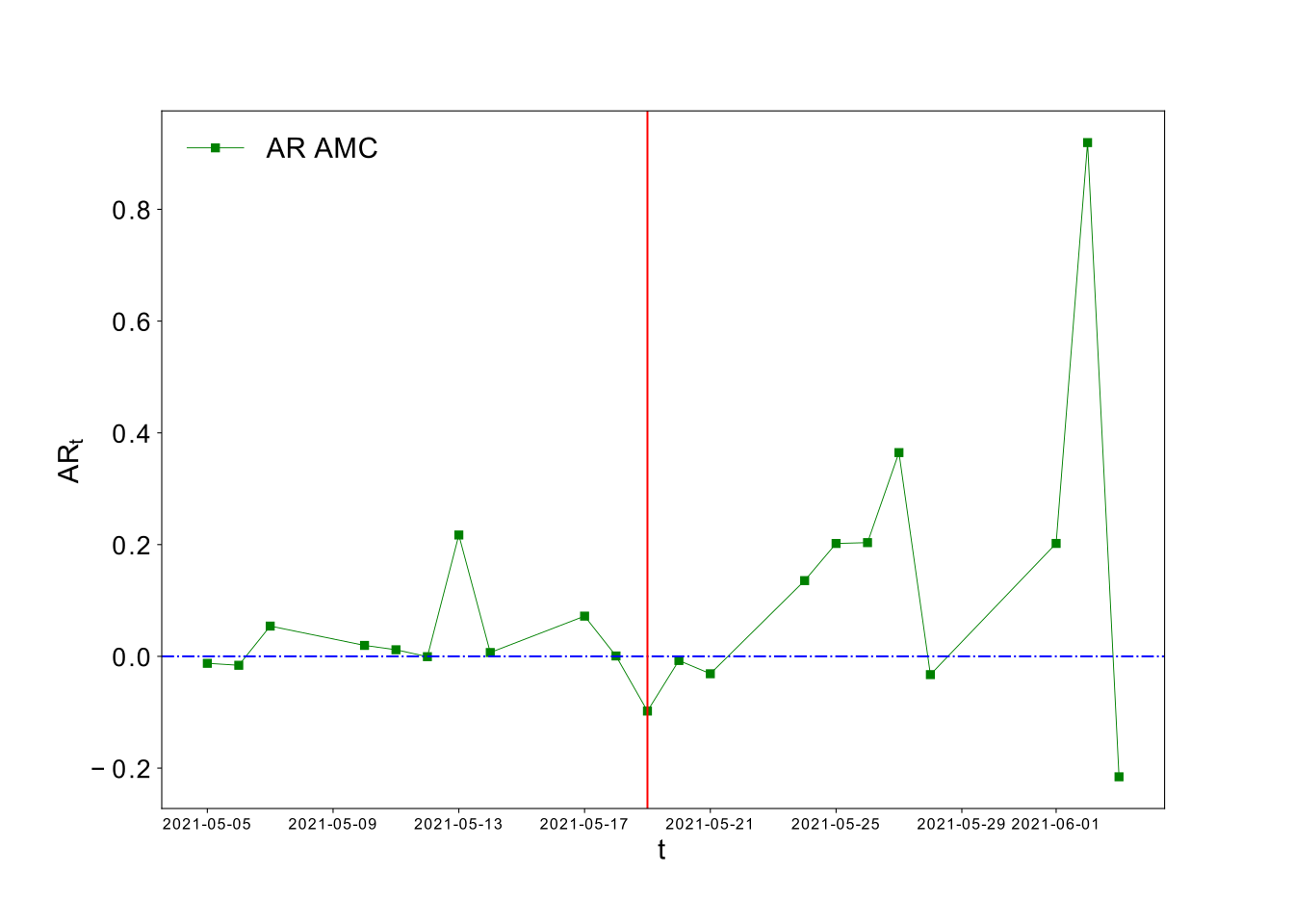}
        \end{subfigure}
        \begin{subfigure}[b]{0.49\textwidth}   
            \centering 
            \includegraphics[width=\textwidth]{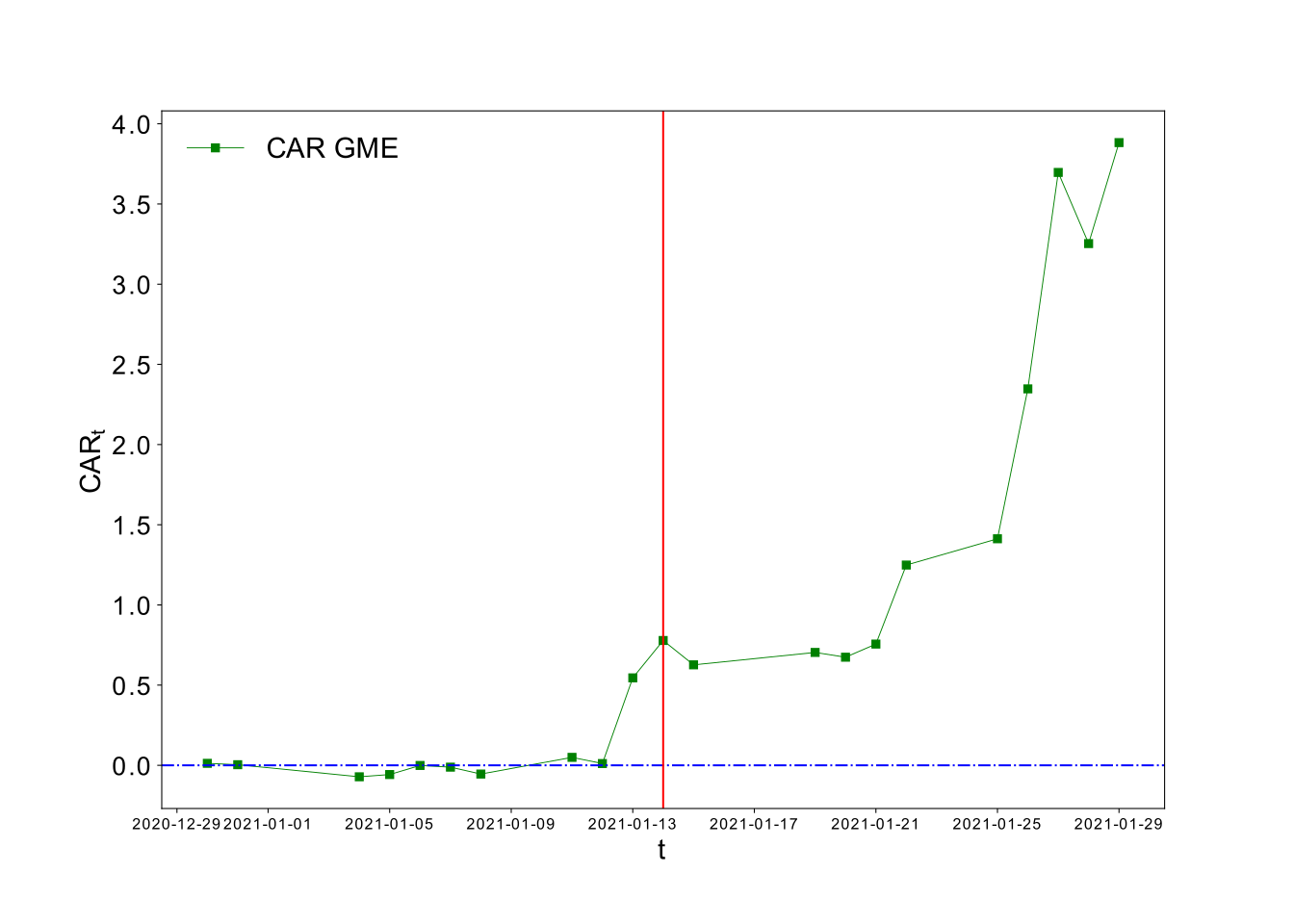}
        \end{subfigure}
        \hfill
        \begin{subfigure}[b]{0.49\textwidth}   
            \centering 
            \includegraphics[width=\textwidth]{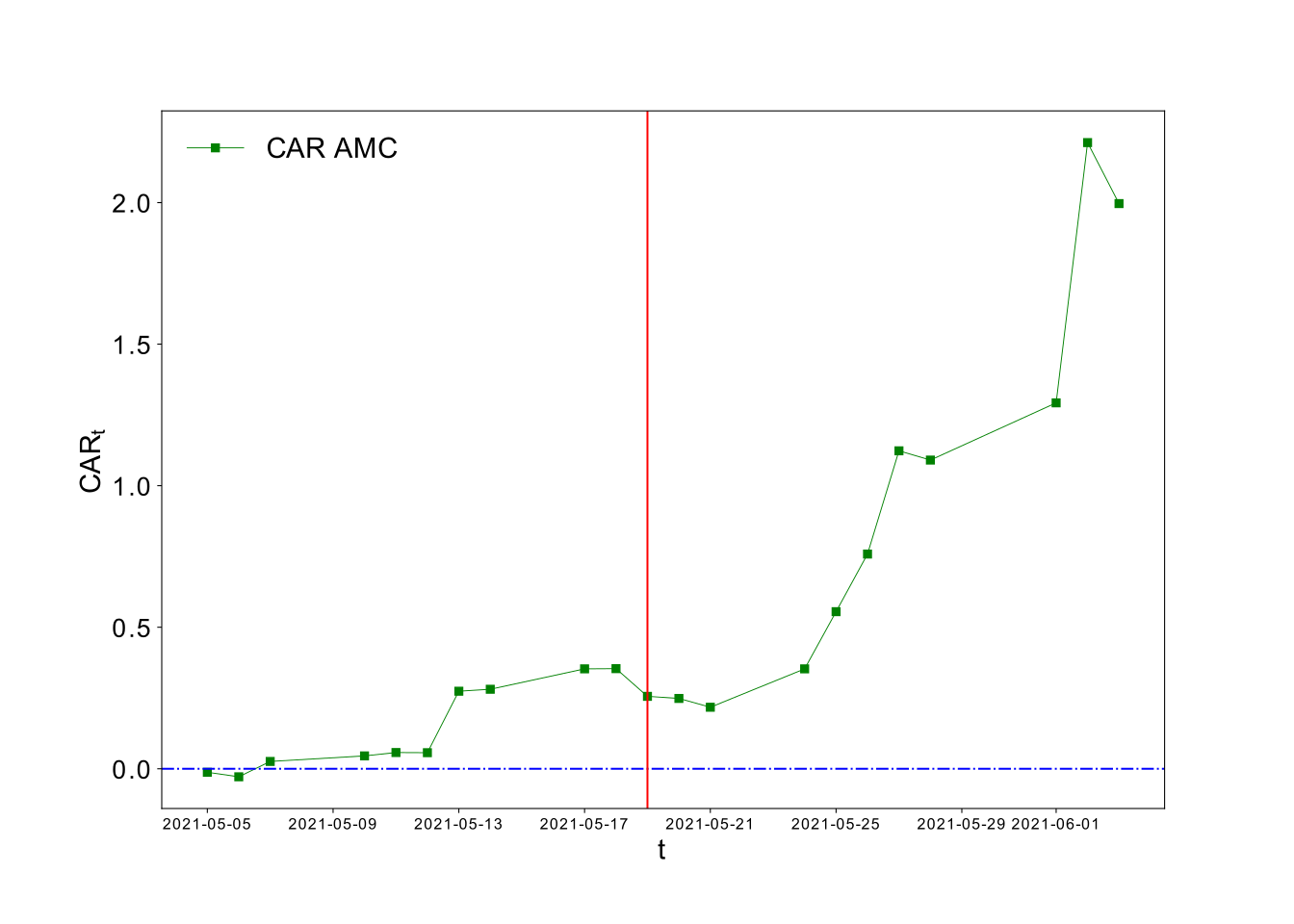}
        \end{subfigure}
        \caption{The figure shows the abnormal returns and the cumulative abnormal returns on a [-10:10] days event window centered in $\tau=0$ that corresponds to the event date, marked with the red vertical line. The event date for GME is January 14th, 2021. The event date for AMC is May 19th, 2021. The horizontal dashed line indicates when the (cumulative) abnormal return equals 0. $AR_t$ and $CAR_t$ plot for GME and AMC single event triggering the highest (cumulative) abnormal returns in the following 10 days.}
        {\small } 
        \label{ar_car_plots}
    \end{figure}

\begin{table}[H]
\centering
\setlength{\tabcolsep}{10pt}
\begin{tabular}{cccc} \toprule
{\hspace{60pt}} &   \\ 
 Day ($\tau$) & $AR_{AMC,t}$   &    $CAR_{AMC,t}$  & $AVol_{AMC,t}$ \\ 
\midrule

-10 & -0.012 & -0.012 & 0.710 \\

-9  & -0.016 & -0.028 & 1.069\\

-8 & $0.054^{*}$ & 0.026 & 0.982\\

-7 & 0.020 & 0.045 & 1.067 \\

-6 & 0.012 & 0.057 &1.156 \\

-5  & -0.0005 & 0.057 & 1.267\\

-4 & $0.217^{**}$ & 0.274 & 6.972\\

-3 & 0.007 & 0.281 & 5.024\\

-2 & $0.072^{*}$ &  0.353 &4.001\\

-1  & 0.0007 & 0.354 &4.438 \\

0 & $-0.098$ & 0.256  &2.277 \\

+1 & -0.007 & 0.248  & 1.569\\

+2 & -0.031 & 0.217 &1.329\\

+3  & $0.135^{**}$ & 0.353  &2.866 \\

+4 & $0.202^{**}$ & 0.555  &5.241\\

+5 & $0.203^{**}$ & 0.758  &9.941\\

+6 & $0.365^{**}$ & 1.123  &\textbf{18.382} \\

+7 & -0.033 & 1.090 &10.702\\

+8  & 0.202$^{**}$ & 1.293 &6.592\\

+9 & \textbf{0.919}$^{**}$ & \textbf{2.212} &8.603\\

+10 & -0.215 & 1.996 & 5.868\\

\bottomrule

\end{tabular}
\caption{Event study for AMC on the event date May 19th, 2021 ($\tau = 0$). This table shows the abnormal returns, the cumulative abnormal returns and the abnormal volumes on a [-10:+10] days event window around the event day. $^{***}$, $^{**}$, $^{*}$ represents significance at 1\%, 5\% and 10\% for the abnormal returns only. The highest value for each column in bold.}
\label{table:ar_amc}
\end{table}
Table \ref{table:ar_gme} and \ref{table:ar_amc} represents the abnormal returns, the cumulative abnormal returns and the abnormal volumes for the stocks GME and AMC over the event window [-10:10]. The event window presents significant abnormal returns for both stocks, but the significance is generally higher for GME and more concentrated after the event. Interestingly, in both stocks, we find a clear upward trend of the abnormal returns after the event, perfectly in line with the assumptions that users are coordinated on the Reddit community to buy the stock massively. Figure \ref{ar_car_plots} corroborates the claim.

\begin{figure}[ht]
    \centering
    \begin{subfigure}[b]{\textwidth}
    \includegraphics[width=0.9\textwidth]{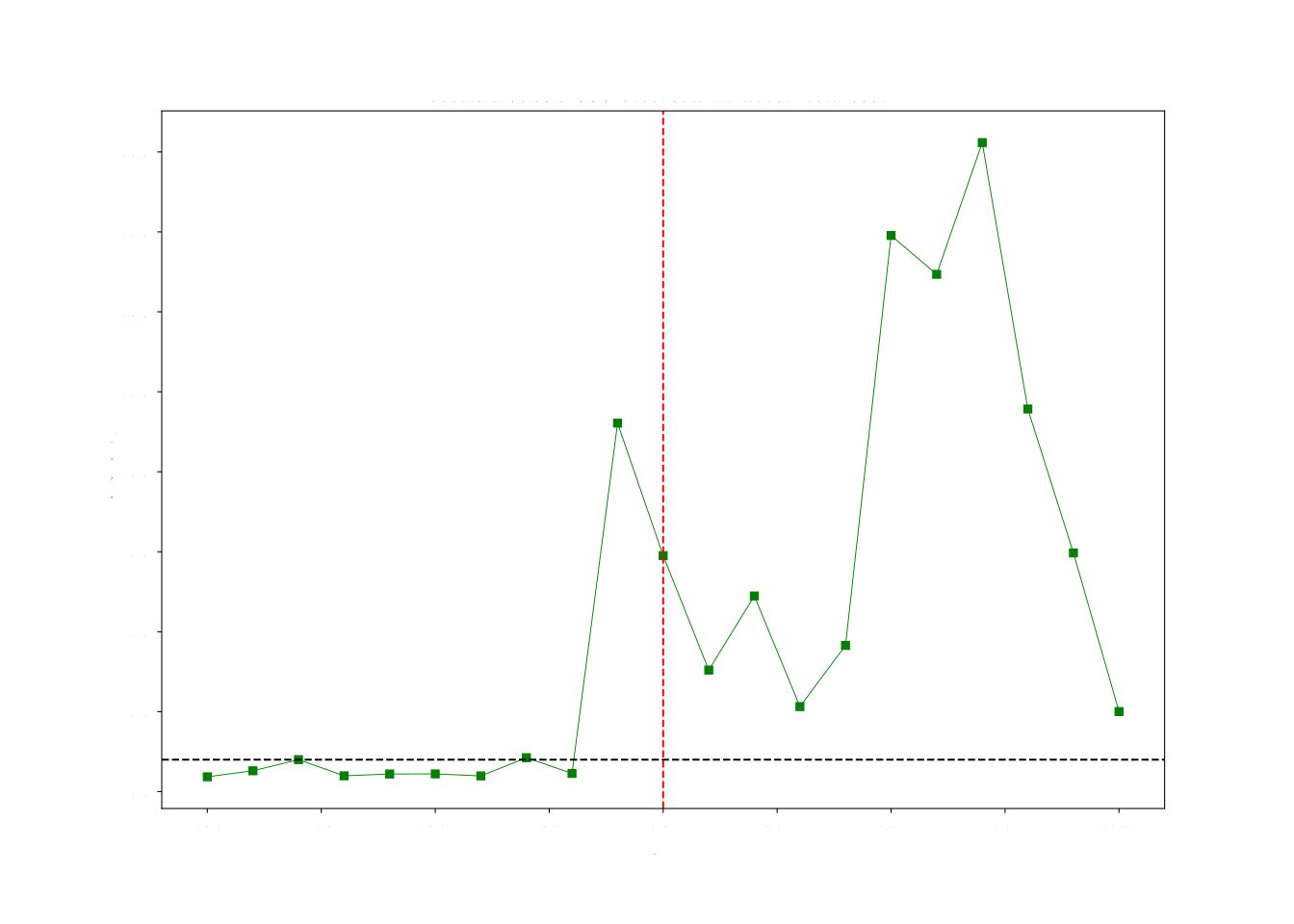}
    \end{subfigure}
    \begin{subfigure}[b]{\textwidth}
    \includegraphics[width=0.9\textwidth]{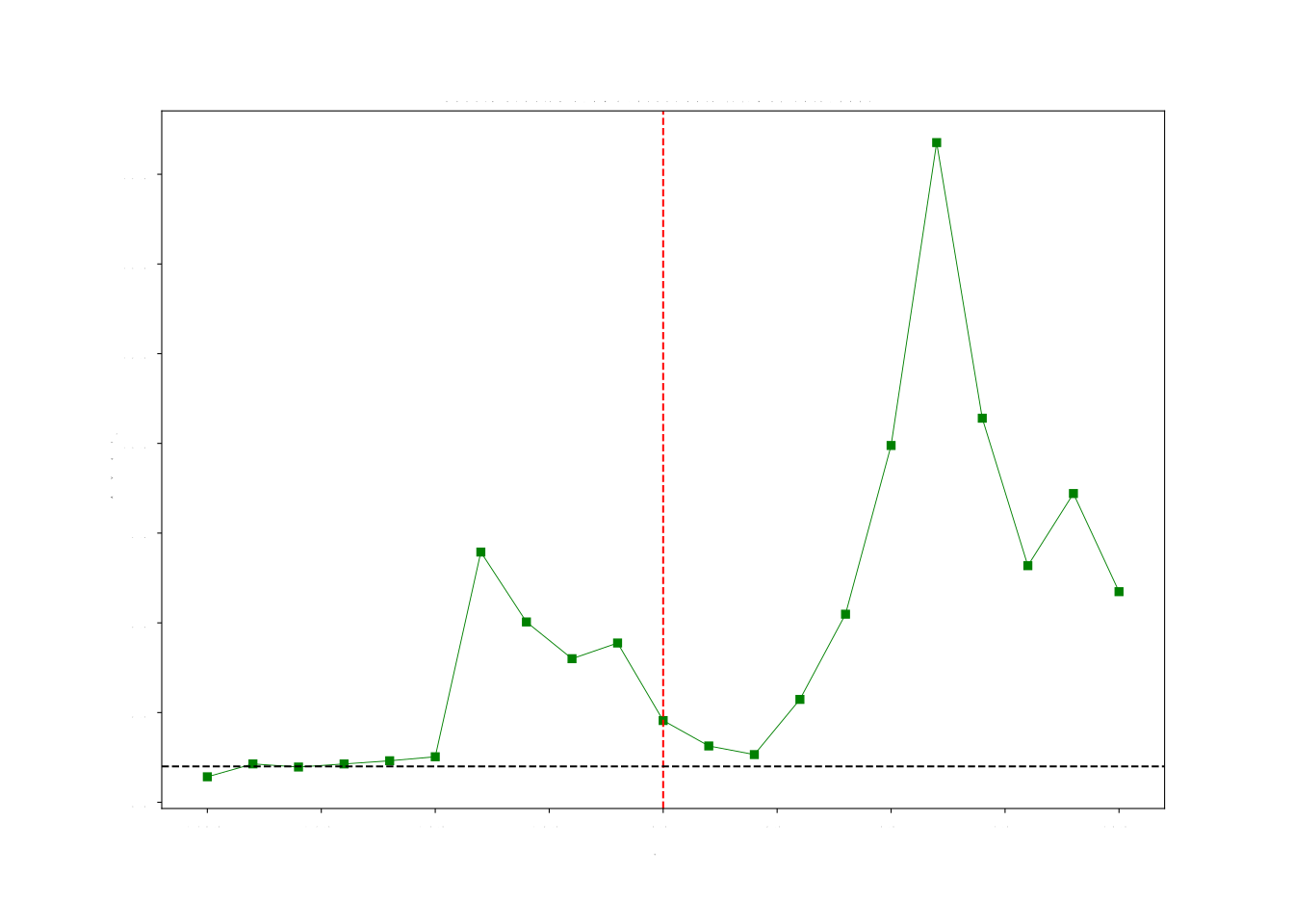}
    \end{subfigure}
    \caption{The figure shows the abnormal volume on a [-10:10] days event window centered in $\tau=0$ that corresponds to the event date, marked with the red vertical dashed line. The event date for GME is January 14th, 2021. The event date for AMC is May 19th, 2021. The horizontal dashed line indicates when the abnormal volume equals 1.}
    \label{fig:Abn_Volume_GME}
\end{figure}

In Figure \ref{fig:Abn_Volume_GME} we report the graphical representation of the event analysis for the abnormal volume (last column of Tables \ref{table:ar_gme} and \ref{table:ar_amc}). In line with the analysis of (cumulative) abnormal returns, the abnormal volume has a similar reaction after the event happens. It presents a significant upward trend after detecting unusual and cooperative activity on the social network.\\
Overall, the event analysis suggests significant abnormal returns following the alert dates for the meme-stock only, GME and AMC. The alert for non-meme stocks (AAPL and MSFT) turns on rarely and without generating abnormal returns. 
The alert results mark significant differences between the meme and non-meme stocks. When the alert system turns on for the meme stocks, it is related to an attempt by retail traders to coordinate an action to drive the stock market price up. The alert system on the non-meme stock detects agents on the social network discussing the stock's performance without necessarily coordinating moves onto the market.

\subsection{Regression analysis}

In this section we focus on the differences between the meme and non-meme stocks' abnormal returns. In particular, we assess the dependence between abnormal returns and social network-retrieved information, finding different patterns for the two stocks' categories. 
We run a time-series regression of abnormal stock returns on social network-related variables, controlling for market-related and stock-specific variables.\\
The contemporaneous regression is estimated via OLS with HAC estimator for the variance by \cite{newey1987hypothesis} for each stock $s$:
\begin{equation}\label{eq:reg}
    AR_{t,s} = \alpha_{s} + Vol_{t,s} + P^{O}_{t,s} + AR_{t-1,s} + Rep_{t} + SubRank_{t} + ABN_{t,s} + O_{t,RH} + ABN_{t,s}\times O_{t,RH} + \varepsilon_{t,s}
\end{equation}
Stock-specific financial variables are the transaction volume $Vol_{t,s}$ and the stock open-price $P^{O}_{t,s}$; they enter the regression in the first difference to mitigate non-stationarity and trends. The lagged term for the abnormal returns $AR_{t-1,s}$ is included.\\
We also include some subreddit specific variables: the variable $Rep_t$ is the daily number of reports of non-functioning of the RobinHood trading application, which is very popular among young retail investors. The data is downloaded from DownDetector.com, an online platform that provides users with real-time information on various websites and services.\\
$SubRank_t$ is a metric computed internally by Reddit to compare the popularity across the various communities as a function of its users. For each community, they compute the number of subscribers and their popularity (according to the karma, the number of upvotes,..) and they rank the subreddits according to this indicator. Hence, the lower the metric, the higher is the popularity of the subreddit. See Appendix \ref{Subscriber_rank}. \\
$O_{t,RH}$ is a dummy assuming a value of 1 when the website DownDetector.com reports the presence of malfunctioning of RobinHood. According to DownDetector.com, malfunctioning is related to the platform's outages or generic problems in the service (like Robinhood being slow in placing trades' orders)\footnote{$O_{t,RH}$ assumes value of 1 when the variable $Rep_t$ is different from 0.}.\\
The stock-specific social-network-related variable $ABN_{t,s}$ is the average branching number, a proxy of the daily virality of the comment-trees. The proxy of virality is inspired by \cite{ChoiEtAl2015} where conversations' thread are defined as tree $T^k_{t,s}$. The average branching number (virality) quantifies on average how many comments a node generates. It is computed as the average number of children at each (non-leaf) node: number of non-root nodes (the volume of the tree minus one $M^{k}_{t,s}-1$, namely the number of edges) divided by the number of non-leaf nodes (the number of nodes with children). In the equation (\ref{eq:reg}) the average branching number interacts with the dummy $O_{t,RH}$.\\
\begin{table}[H]\centering
\newcommand\sym[1]{\rlap{$^{#1}$}}

\begin{tabular*}{\columnwidth}{
  @{\hspace{\tabcolsep}\extracolsep{\fill}}
  l*{3}{{c}{c}{c}}
}

\toprule
  & \multicolumn{2}{c}{Meme stock} & \multicolumn{1}{c}{} & \multicolumn{2}{c}{Non meme stock}\\ \cline{2-3} \cline{5-6}
  &\multicolumn{1}{c}{GME}&\multicolumn{1}{c}{AMC}&\multicolumn{1}{c}{}&\multicolumn{1}{c}{AAPL}&\multicolumn{1}{c}{MSFT}\\
\midrule
\textit{$\alpha_{t,s}$} & 0.0262  & -0.0267 &  & 0.0021 &  0.0007   \\
        & (0.657)  & (-1.559 )  & & (1.464)  & (0.485)   \\
\addlinespace
\textit{Vol$_{t,s}$} & 0.0609$^{***}$   & 0.1791$^{***}$   & & -0.0033$^{**}$  & -0.0041$^{***}$ \\
        & (3.026)  & (4.358)  & & (-1.976)  & (-3.573)   \\
\addlinespace
\textit{$P^{O}_{t,s}$} & 0.1162$^{***}$  &  0.0469$^{*}$  & & 0.0099$^{***}$  & 0.0066$^{***}$ \\
        & (5.254)  & (1.878) & & (6.361)  & (3.408)   \\
\addlinespace     
\textit{$AR_{t-1,s}$} &  -0.0597 & 0.0600  & & -0.3247$^{***}$  & -0.2589$^{**}$\\
        & (-0.551)  & (0.878) &  & (-3.320)  & (-2.126)   \\ 
\addlinespace     
\textit{$Rep_{t}$} & 0.5224$^{**}$  & 0.5826$^{***}$& &0.0026 & 0.0135$^{***}$  \\
        & (2.260)  & (2.659) &  & (0.363) & (2.718)    \\ 
\addlinespace     
\textit{$SubRank_{t}$} & 0.0168$^{**}$  & 0.0204$^{**}$ &  & -0.0019  & -0.0019$^{*}$ \\
        & (2.121)  & (2.009)  & & (-1.493) & (-1.860)    \\    
\textit{$ABN_{t,s}$} & -0.0161  & 0.0026 &  &  -0.0002 & 0.0008  \\
        & (-1.067)  & (0.423) & &(-0.268) &    (0.674)  \\
\addlinespace        
\textit{$O_{t,RH}$} & -0.2097$^{**}$ & -0.0815&  & -0.0164$^{***}$ & -0.0003 \\
        &  (-2.171) &  (-1.648) & & (-3.834)  & (-0.099)   \\          
\textit{$ABN_{t,s} \times O_{t,RH}$} & 0.0738$^{**}$  & 0.0632$^{**}$ & & -0.0513$^{***}$ & -0.0013    \\
        & (2.000) & (2.333) & &(-3.313) & (-0.791)  \\

\addlinespace        
\addlinespace

\addlinespace

\midrule
Observations: & \multicolumn{1}{c}{166} & \multicolumn{1}{c}{166}& \multicolumn{1}{c}{} & \multicolumn{1}{c}{166}& \multicolumn{1}{c}{166} \\
Adjusted $R^2$: & \multicolumn{1}{c}{0.459} & \multicolumn{1}{c}{0.618} & \multicolumn{1}{c}{} & \multicolumn{1}{c}{0.192}&\multicolumn{1}{c}{0.190} \\
AIC: & \multicolumn{1}{c}{-118.5} & \multicolumn{1}{c}{-92.07} & \multicolumn{1}{c}{} & \multicolumn{1}{c}{-884.6} & \multicolumn{1}{c}{-947.6}\\

\bottomrule
\multicolumn{4}{l}{\footnotesize \textit{t} statistics in parentheses}\\
\multicolumn{4}{l}{\footnotesize * \(p<0.05\), ** \(p<0.01\), *** \(p<0.001\)}\\
\multicolumn{4}{l}{\footnotesize}\\
\end{tabular*}
\caption{Contemporaneous regression of daily abnormal return. In each column the dependent variable is the abnormal return $AR_{t,s}$ and the independent variables are stock or subreddit specific characteristics. The regression is estimated via OLS with HAC estimator for the standard errors. \label{tab:reg_gme}}
\end{table}

Table \ref{tab:reg_gme} summarizes regression results. The variable $Subrank_t$ presents significant and positive coefficients for the meme-stock, while it is negative and significant only for MSFT. An increase in the average rank of users on the Wallstreetbets community positively impacts the abnormal returns, meaning that the prestige of the users relates to a positive increase in the abnormal returns of the meme-stocks. The same does not happen for the non-meme stocks. The number of Robinhood outage reports has a positive and significant effect on the abnormal returns for the meme-stock. The effect of reports for AAPL is not significant, while for MSFT is still positive and significant, but lower in magnitude than the meme-stocks. The increase in the number of reports may be related to high activity in the stock market for buying stock discussed on the Wallstreetbets community. This effect is associated with an increase in abnormal returns, and it is even more marked for the meme-stock.
Interestingly, the daily virality of the comment trees positively impacts the abnormal returns during the day of outages of the Robinhood platform.
While interacting the average branching number with outages positively impacts abnormal returns for GME, the dummy outages alone is associated with negative abnormal returns. Intuitively, this might be related to big players' moves on the GME stock. During the Robinhood outages, banks and hedge funds operated in the market without the noisy traders that usually invest through the Robinhood platform. Their action is associated with a fall in price. Indeed, during the day of the GME squeeze - where the price peaked and started to fall vertiginously - the Robinhood platform was shut down. As the average branching number (virality) is related to the higher activity of the users on the Wallstreetbets community, considering it during the platform outages, may disentangle the effect of noisy traders rather than big corporations on the stock. The price of the stock increase during days of high virality and platform outages, which might be related to the noisy traders' action.
The coefficient on the transaction volume is negative for the non-meme stock, and in line with the standard liquidity theory\footnote{Liquidity is the ease of trading securities. It is the degree to which a significant quantity of an asset can be traded within a short time frame without incurring a significant transaction cost or adverse price impact.}: a higher trading volume commands a lower expected stock return because liquidity is a desired stock feature for a risk-averse agent\footnote{Numerous works are confirming this evidence. For instance, \cite{AcharyaPedersen2005} and \cite{ChordiaEtAl2001}. }. For the meme-stock case, the transaction volume has an unexpected significant positive effect on abnormal returns\footnote{We run a predictive regression on the abnormal returns, presenting the results in the appendix. The purpose of the predictive regression is to evaluate whether social network-retrieved information adds information when market-related variables, such as transaction volume and price, are included in the regression.}.\\
To evaluate the echo chamber effects from Reddit to the stock market, we run a regression with the abnormal volume as a dependent variable:
To evaluate the role of users in trading the meme and non-meme stocks, we run a regression where the dependent variable is the abnormal volume:
\begin{equation}\label{eq:reg_xxx}
    AV_{t,s} = \alpha_{s} + Vol_{t,s} + P^{C}_{t,s} + AV_{t-1,s} + Sub_{t} + O_{t,RH} + Sub_{t}\times O_{t,RH}
\end{equation}
The close price $P^{C}_{t,s}$ is used as a regressor. The variable $Sub_t$ represents the total number of subscribers to the Wallstreetbets community each day, and interacts with the dummy outages.
\begin{table}[H]\centering
\newcommand\sym[1]{\rlap{$^{#1}$}}
\begin{tabular*}{\columnwidth}{
  @{\hspace{\tabcolsep}\extracolsep{\fill}}
  l*{3}{{c}{c}{c}}
}
\toprule
& \multicolumn{2}{c}{Meme stock} &  &\multicolumn{2}{c}{Non-meme stock}\\\cline{2-3} \cline{5-6}
  &\multicolumn{1}{c}{GME}&\multicolumn{1}{c}{AMC}& \multicolumn{1}{c}{}&\multicolumn{1}{c}{AAPL}&\multicolumn{1}{c}{MSFT}\\
\midrule
\textit{$\alpha_{t,s}$} & 0.1862$^{**}$  & 0.3565$^{***}$   & & 0.0321$^{***}$ &  0.0373$^{***}$   \\
        & (2.539)  & (3.058)  & & (2.926)  & (2.661)   \\
\addlinespace
\textit{Vol$_{t,s}$} & 1.0646$^{***}$   & 0.2971$^{***}$   & & -0.0033$^{**}$  & 0.2858$^{***}$ \\
        & (4.1555)  & (8.350) &  & (45.209)  & (63.264)   \\
\addlinespace
\textit{$P^{C}_{t,s}$} & 0.0806  &  -0.1437$^{*}$ & & 0.0038  &  -0.0004 \\
        & (1.412)  & (-1.879) & & (0.956)  & (-0.137)   \\
\addlinespace     
\textit{$AV_{t-1,s}$} &  0.7315$^{***}$ & 0.0600  & & 0.9596$^{***}$  & 0.9618$^{***}$\\
        & (12.025)  & (9.396)  & & (73.303)  & (63.706)   \\ 
\addlinespace        
\textit{$Sub_t$} & 0.0219 & -0.0109 & & -0.0023 & -0.0010 \\
        &  (0.464) &  (-0.120) &  & (-0.618)  & (-0.345)   \\   
\textit{$O_{t,RH}$} & 0.4222$^{*}$ & 0.1663 &  & 0.0062 & 0.0028 \\
        &  (1.875) &  (1.016)&  & (0.906)  & (0.422)   \\   
\textit{$Sub_t \times O_{t,RH}$} & -0.3487$^{*}$  & -0.2621  & & 0.0038 & -0.0001    \\
        & (-1.908) & (-1.483)& & (0.542) & (-0.017)  \\

\addlinespace        
\addlinespace

\addlinespace

\midrule
Observations: & \multicolumn{1}{c}{166} & \multicolumn{1}{c}{166} & \multicolumn{1}{c}{} & \multicolumn{1}{c}{166}& \multicolumn{1}{c}{166} \\
Adjusted $R^2$: & \multicolumn{1}{c}{0.759} & \multicolumn{1}{c}{0.731} & \multicolumn{1}{c}{} & \multicolumn{1}{c}{0.977}&\multicolumn{1}{c}{0.989} \\
\bottomrule

\multicolumn{4}{l}{\footnotesize \textit{t} statistics in parentheses}\\
\multicolumn{4}{l}{\footnotesize * \(p<0.05\), ** \(p<0.01\), *** \(p<0.001\)}\\
\multicolumn{4}{l}{\footnotesize}\\
\end{tabular*}
\caption{Contemporaneous regression of daily abnormal volume. In each column the dependent variable is the abnormal volume $AV_{t,s}$ and the independent variables are stock or subreddit specific characteristics. The regression is estimated via OLS with HAC estimator for the standard errors.\label{tab:reg_gme_av}}
\end{table}

Interestingly, the market variables are highly significant for the non-meme stock (mainly because they are less volatile and less driven by social networks-related variables), pushing towards the adjusted $R^2$. 
The striking part of Table 4 is the first column, where the dependent variable is the abnormal volume for GME. The coefficient on outages is positive: this means that big corporations invested in the stock during the Robinhood shutdown. The effect on the price was negative (Table 3): this led us to think that noisy traders on Reddit essentially move the price forward by massively buying the stock. At the same time, big banks and hedge funds drove the fall in price on January 29th (short-squeeze) when the platform was shut down. The number of subscribers during the Robinhood outages negatively impacts the abnormal volume; therefore, higher activity on the Wallstreetbets community is not associated with higher transaction volumes when Robinhood is shut down. This confirms that investors are inexperienced and noisy traders who mainly use the platform Robinhood to buy the stock.

\section{Final discussion}
Social media are a powerful and impacting tool to disseminate information and stir a vast mass of people. This paper provides empirical evidence of the echo chamber effect on the financial market from the social network, a form of market manipulation (albeit indirect). 
Reddit and, in general, social media are great places to share advice and manipulate poorly-informed, unsophisticated, and prone to be convinced investors.\\
We provide evidence that the manipulators, or to use a word a là Pedersen, the fanatics can coordinate many naive investors to provoke the desired stock price movement. The fanatics can effectively undermine the financial market stability by persuading inexperienced and easily reachable people.\\
We design an alert system to detect abnormal movement related to a specific stock on social media based on the extraordinary activity in terms of volume and the detection of a potential manipulator that coordinates the mass movement.\\
While it is far from our duty to evaluate whether the promotion and persuasion practice falls within the boundaries of the law, our consideration concerns the market microstructure models. In front of these episodes, the retail investors can no longer be relegated to a residual category of 'noise traders', but models should consider that many small and apparently harmless investors if aggregated and coordinated, can generate a disruptive effect on the financial market.\\
In the end, the entire analysis spots significant differences between the meme and non-meme stocks. The detection system finds alerts for both categories. Still, they never turn into abnormal returns for the non-meme stocks, suggesting fewer chances of suspicious trading activity or market manipulation in those cases. Moreover, the regression analysis indicates that social network-retrieved information is significant for meme stock abnormal returns only, resulting in structural differences between the price formation of the two categories. Indeed, noisy traders may determine the price of meme stocks through social activity and potential coordination.\\
\newpage
\appendix
\appendixpage
\begin{appendices}
\section{Data download}\label{Appendix_data_download}
For each tree, we have as many rows in the data frame as the number of comments, and each row contains the following information:
\begin{itemize}
    \item \texttt{title}: the textual content of the initial submission;
    \item \texttt{body}: the textual content of the comment;
    \item \texttt{name}: the id of the author of the comment (each id is prefixed by \texttt{'t1\_'} to specify the author made a comment activity);
    \item \texttt{parent\_id}: the author of the parent comment to which the comment in question refers to (the \texttt{parent\_id} can be prefixed by \texttt{'t1\_'} if the author of the comment replies to an other comment or it can be prefixed by \texttt{'t3\_'} if the author of the comment replies to the top-level post, i.e. the submission);
    \item \texttt{author\_name}: the name of the author who post the initial submission;
    \item \texttt{depth}: the level of the comment tree at which the comment in question is located (if a tree is composed by the initial submission only, the depth is 0; if the comment refers to the initial submission the depth is 1; if the comment refers to a comment in the first level, the depth is 2; and so on);
    \item \texttt{score}: the number of up-votes minus the number of down-votes obtained by the comment;
    \item \texttt{score\_submission}: the number of up-votes minus the number of down-votes obtained by the initial submission;
    \item \texttt{upvote\_ratio}: the percentage of upvotes on the total votes received by a submission;
    \item \texttt{time\_submission}: date and time at which the initial submission is published;
    \item \texttt{time\_comment}: date and time at which the comment is published;
    \item \texttt{num\_comment}: number of comments below the initial submission that compose the tree;
    \item \texttt{flair}: a tag used to categorize the post according to the topic it deals with; they are subreddit specific and in the case of the subreddit \textit{r/WallStreetBets} the users can select among the following ones:
    \begin{itemize}
        \item YOLO, the acronym for 'You Only Live Once, it can be used for posts presenting extremely aggressive investment strategies with a consistent value at risk;
        \item DD, the acronym for Due Diligence, must be applied to post presenting research on a specific company/sector/trade. It should include sources and citations;
        \item Discussion, an idea or article that you would like to talk about;
        \item Gain, to show off a solid winning trade;
        \item Loss, to show off a brutal, crushing loss;
        \item Earnings Thread, weekly earnings discussion thread or a specific earnings event;
        \item Daily Discussion, daily catch-all thread for discussions;
        \item Mods, only for official business.
    \end{itemize}
    \item \texttt{distinguished}: if a bot automatically performs the commenting activity, the variable reports the wording 'Moderator', none otherwise, when a non-automatic user adds the comment.
\end{itemize}
Note that in the case of submission without comments below, the data frame has a single row with empty values for the comments-related variables.\\
\newpage
\section{The Social Network of Reddit users}\label{Network_graph}
\begin{figure}[ht]
    \centering
    \includegraphics[width=\textwidth]{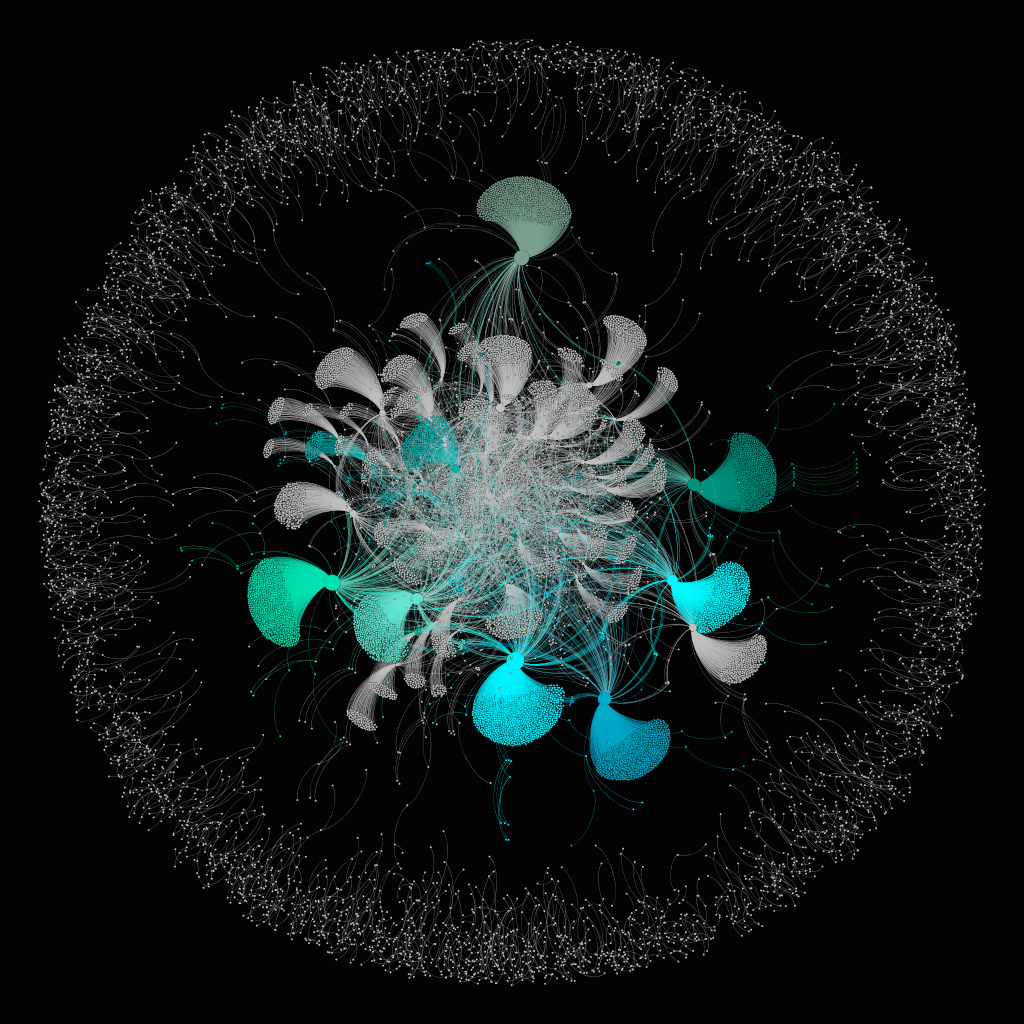}
    \caption{The Figure shows the network of users interacting on Reddit on January 31st, 2021. The network reports the interactions of users posting a submission containing the wording 'AMC'. The graph contains 15.534 nodes and 21.032 edges. The colored part of the network is the nodes involved in the net of the 10 users with the highest in-degree centrality.}
    \label{fig:Network_graph_AMC_31_01_21}
\end{figure}

\begin{figure}[ht]
    \centering
    \begin{subfigure}[t]{0.4\textwidth}
    \centering
    \includegraphics[width=\textwidth]{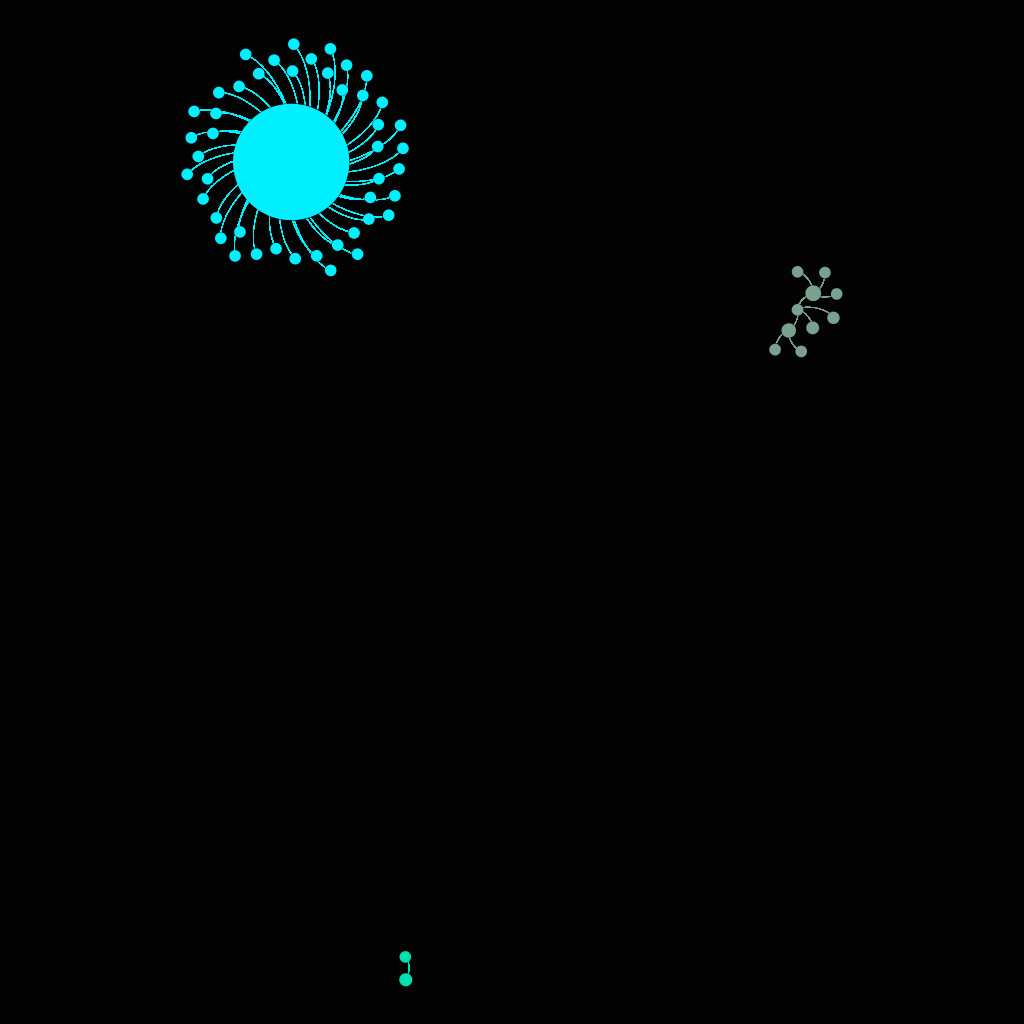}
    \end{subfigure}
    \begin{subfigure}[t]{0.4\textwidth}
    \centering
    \includegraphics[width=\textwidth]{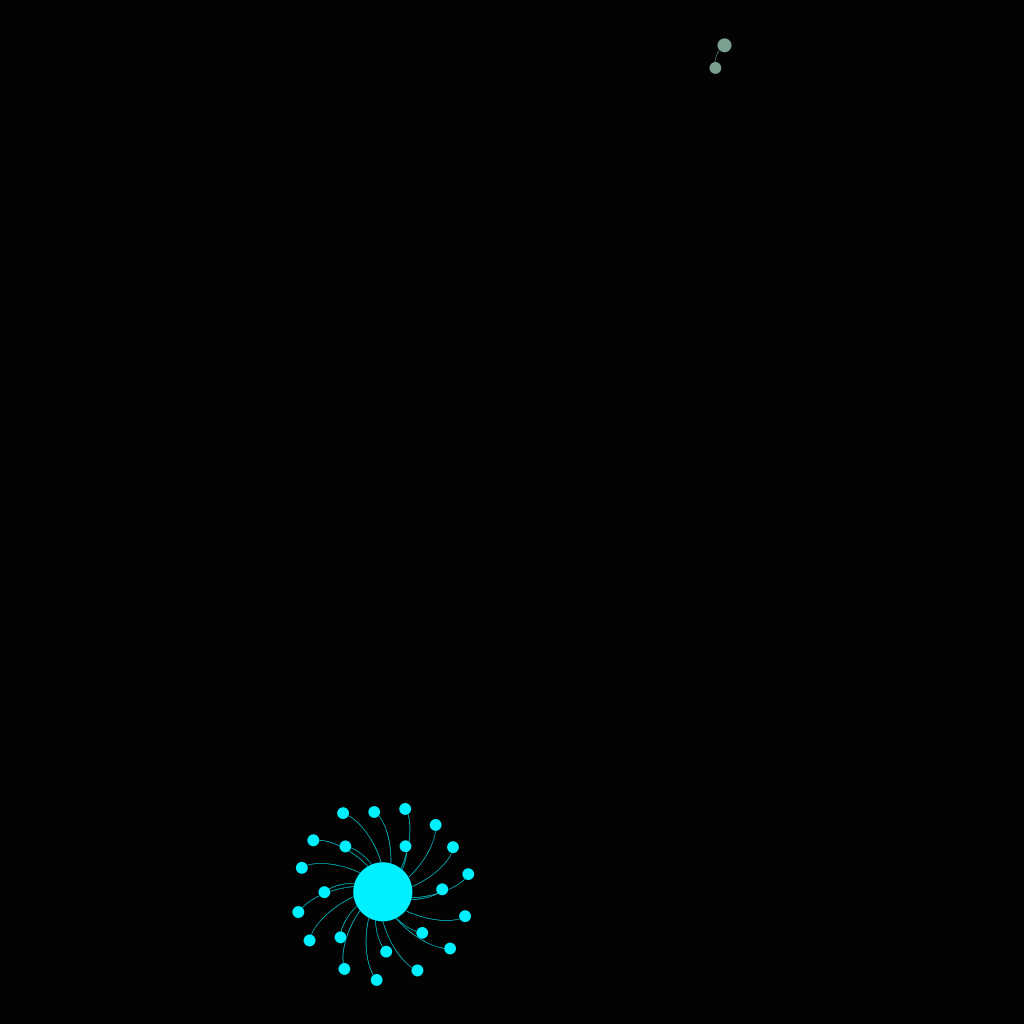}
    \end{subfigure}
    \caption{The Figure shows the network of users interacting on Reddit on June 22nd, 2021 for AAPL and on May 20th, 2021 for MSFT. The network shows the interactions of users posting a submission containing the wording 'AAPL' and 'MSFT' respectively. }
    \label{fig:network_graph_AAPL_MSFT}
\end{figure}

In Figure \ref{fig:Network_graph_AMC_31_01_21}, we present an example of network user graph on January 31st, 2021, where the main submissions contain the ticker AMC. There are 15.534 users (represented by the nodes) interacting among them on the platform throughout commenting activity (the 21.032 edges connecting them).\\
The directed edges point from the comment's author to the author of the main submission. The peripheral nodes in the graph are the less connected users; in the central part of the network, the most connected and central users: the colored ones are the users with the highest in-degree centrality.\\ 
Figure \ref{fig:network_graph_AAPL_MSFT} presents the network graph for AAPL a and MSFT on two alert dates, respectively June 22nd, 2021 for AAPL and on May 20th, 2021 for MSFT. Compared to the meme-stock case, the non-meme-stocks present a feeble activity on the social network even in extraordinary occasions that determine the alert activation.
\newpage
\section{Subscriber rank}\label{Subscriber_rank}
\begin{figure}[ht]
    \centering
    \includegraphics[width=0.9\textwidth]{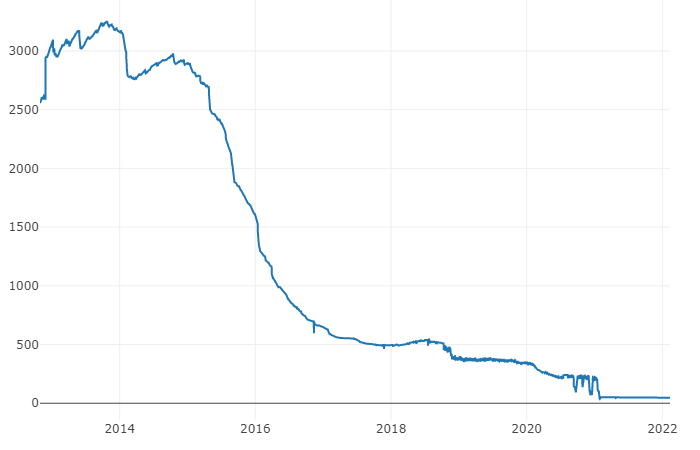}
    \caption{The figure shows the time series of the indicator subscriber rank in the period 2013-2022.}
    \label{fig:subrank}
\end{figure}

In Figure \ref{fig:subrank} we represent how the Reddit indicator 'Subscriber rank' varies over time, in the period 2013-2022.

\section{Predictive regression}
This section aims at understanding whether social network information enhances the predictability of abnormal returns. We run a regression where future abnormal returns depend on market-related variables and variables from the social network.
\begin{equation}\label{eq:reg_pr}
    AR_{t,s} = \alpha_{s} + Vol_{t-1,s} + Vol_{t-2,s} + P^{O}_{t-1,s} + P^{O}_{t-2,s}  + rank_{t-1} + rank_{t-2} + Rep_{t} + \varepsilon_{t,s}
\end{equation}
The variable $rank_{t}$ represents the daily average rank between subscribers and users commenting on submissions. 
\begin{table}[H]\centering
\newcommand\sym[1]{\rlap{$^{#1}$}}

\begin{tabular*}{\columnwidth}{
  @{\hspace{\tabcolsep}\extracolsep{\fill}}
  l*{2}{{c}{c}{c}}
}
\toprule
 & \multicolumn{2}{c}{Meme stock}\\ \cline{2-3}
  &\multicolumn{1}{c}{GME}&\multicolumn{1}{c}{AMC} \\ 
\midrule
\textit{$\alpha_{t,s}$} & -0.0825$^{*}$ & -0.0942$^{**}$   \\
        & (-1.705) & (-2.602)  \\
\addlinespace
\textit{Vol$_{t-1,s}$} & 0.0350$^{***}$ & -0.0401 &  & \\
        & (2.828) & (-1.063)    \\
\textit{Vol$_{t-2,s}$} & 0.0157 & 0.0038  \\
        & (1.004) & (0.168)   \\
\textit{$P^{O}_{t-1,s}$} & -0.0341 & -0.0274  \\
        & (-1.321) & (-1.293)   \\
\textit{$P^{O}_{t-2,s}$} & 0.0467$^{**}$ & 0.0212  \\
        & (2.529) & (0.645)   \\
\textit{$rank_{t-1,s}$} & 0.0199$^{**}$ & 0.0252$^{*}$  \\
        & (2.042) & (1.663)    \\
\textit{$rank_{t-2,s}$} & 0.0040 & 0.0019  \\
        & (0.445) & (0.140)    \\
\textit{$Rep_{t,s}$} & 0.6684$^{**}$ & 0.9110$^{*}$  \\
        & (2.290) & (1.697)    \\

\addlinespace        
\addlinespace

\addlinespace

\midrule
Observations: & \multicolumn{1}{c}{164} & \multicolumn{1}{c}{164} \\
Adjusted $R^2$: & \multicolumn{1}{c}{0.243} & \multicolumn{1}{c}{0.211} \\
\bottomrule
\multicolumn{2}{l}{\footnotesize \textit{t} statistics in parentheses}\\
\multicolumn{2}{l}{\footnotesize * \(p<0.05\), ** \(p<0.01\), *** \(p<0.001\)}\\
\multicolumn{2}{l}{\footnotesize}\\
\end{tabular*}
\caption{Predictive regression of daily abnormal return. In each column the dependent variable is the abnormal return $AR_{t,s}$ and the independent variables are the lagged stock or subreddit specific characteristics. The regression is estimated via OLS with HAC estimator for the standard errors.\label{tab:reg_pred_ar}}
\end{table}

We do not include non-meme stock results because they are not significant. The 1-day lagged volume is significant in predicting the abnormal returns of GME, while the 1-day lagged rank well predicts both GME and AMC abnormal returns. The number of reports remains significant when included in the regression with a positive coefficient.
This regression shows how social network-retrieved information to standard market variables such as volume and stock price when predicting abnormal returns.

\end{appendices}
\clearpage
\singlespacing
\setlength\bibsep{3pt}
\bibliographystyle{elsarticle-harv}
\bibliography{bibliography.bib}

\end{document}